\documentclass[10pt,twocolumn,twoside]{IEEEtran}
\usepackage[utf8]{inputenc}
\usepackage[pdftex]{graphicx}
\usepackage[cmex10]{amsmath}
\usepackage{mathtools}
\usepackage[textsize=scriptsize, textwidth=18mm]{todonotes}
\usepackage{comment}
\usepackage{setspace}
\usepackage{tabularx}
\usepackage{booktabs}

\usepackage{cite}	
\makeatletter
	\def\bstctlcite{\@ifnextchar[{\@bstctlcite}{\@bstctlcite[@auxout]}}
	\def\@bstctlcite[#1]#2{\@bsphack
	\@for\@citeb:=#2\do{%
	\edef\@citeb{\expandafter\@firstofone\@citeb}%
	\if@filesw\immediate\write\csname
	#1\endcsname{\string\citation{\@citeb}}\fi}%
	\@esphack}
\makeatother

\usepackage[linesnumbered,algoruled]{algorithm2e}
\SetKwInput{KwParam}{Parameters}
\SetKwInput{KwNotes}{Notes}
\SetKwComment{Comment}{}{}
\SetCommentSty{small}
\makeatletter
\newcommand{\removelatexerror}{\let\@latex@error\@gobble}
\makeatother

\newtheorem{lemma}{Lemma}
\newtheorem{conjecture}{Conjecture}
\newtheorem{remark}{Remark}

\usepackage{pgfplots}
\pgfplotsset{compat=1.13}
\usetikzlibrary{plotmarks}
\newlength\figureheight 
\newlength\figurewidth 
\pgfplotsset{global_axis_style/.style={
	title style={font=\footnotesize},
	legend style={font=\scriptsize},
	legend style={inner xsep=2pt, inner ysep=1pt, nodes={inner sep=0.8pt}},
	legend style={/tikz/every even column/.append style={column sep=2pt}},
	label style={font=\footnotesize},
	xlabel shift=-3pt,
	ylabel shift=-3pt,
	xticklabel style={font=\footnotesize},
	yticklabel style={font=\footnotesize},
	every axis plot/.append style={line width=1pt}
}} 

\usepackage{bm}
\usepackage{xspace}		
\usepackage{amssymb}	
\usepackage{bbold}		

\makeatletter
 \begingroup
 \catcode`\_=\active
 \protected\gdef_{\@ifnextchar|\subtextup\sb}
 \endgroup
 \def\subtextup|#1|{\sb{\textup{#1}}}
 \AtBeginDocument{\catcode`\_=12 \mathcode`\_=32768 }
\makeatother

\newcommand{\dd}{\mathrm{d}}
\newcommand{\proptoe}{\propto^\mathrm{e}}

\newcommand{\ii}{^{-1}}
\newcommand{\db}{\ensuremath{\,\textrm{dB}}\xspace}
\newcommand{\const}{\ensuremath{\mathrm{const.}}}

\newcommand{\h}{^\mathrm{H}}
\newcommand{\T}{^{\mathrm{T}}}

\DeclareMathOperator*{\argmax}{arg\,max}

\DeclareMathOperator{\diag}{diag}

\DeclareMathOperator{\CN}{CN}

\DeclareMathOperator{\unif}{unif}
\DeclareMathOperator{\re}{Re}

\newcommand{\ee}[1]{\exp\!\left(#1\right)}
\newcommand{\E}[1]{\mathbb{E}\!\left[#1\right]}
\DeclarePairedDelimiter{\norm}{\lVert}{\rVert}
\DeclarePairedDelimiter{\ceil}{\lceil}{\rceil}

\newcommand{\bba}{\mathbb{A}}

\newcommand{\bbr}{\mathbb{R}}
\newcommand{\bbc}{\mathbb{C}}

\newcommand{\cala}{\ensuremath{\mathcal{A}}\xspace}

\newcommand{\calc}{\ensuremath{\mathcal{C}}\xspace}
\newcommand{\cald}{\ensuremath{\mathcal{D}}\xspace}

\newcommand{\calk}{\ensuremath{\mathcal{K}}\xspace}
\newcommand{\call}{\ensuremath{\mathcal{L}}\xspace}
\newcommand{\calm}{\ensuremath{\mathcal{M}}\xspace}

\newcommand{\calo}{\ensuremath{\mathcal{O}}\xspace}
\newcommand{\calp}{\ensuremath{\mathcal{P}}\xspace}

\newcommand{\cals}{\ensuremath{\mathcal{S}}\xspace}

\newcommand{\balpha}{\ensuremath{\bm{\alpha}}\xspace}

\newcommand{\bgamma}{\ensuremath{\bm{\gamma}}\xspace}
\newcommand{\bGamma}{\ensuremath{\bm{\Gamma}}\xspace}

\newcommand{\bmu}{\ensuremath{\bm{\mu}}\xspace}

\newcommand{\bsigma}{\ensuremath{\bm{\sigma}}\xspace}
\newcommand{\bSigma}{\ensuremath{\bm{\Sigma}}\xspace}
\newcommand{\btau}{\ensuremath{\bm{\tau}}\xspace}

\newcommand{\bpsi}{\ensuremath{\bm{\psi}}\xspace}
\newcommand{\bPsi}{\ensuremath{\bm{\Psi}}\xspace}

\newcommand{\ba}{\ensuremath{\bm{a}}\xspace}
\newcommand{\bA}{\ensuremath{\mathbf{A}}\xspace}

\newcommand{\bc}{\ensuremath{\bm{c}}\xspace}
\newcommand{\bC}{\ensuremath{\mathbf{C}}\xspace}

\newcommand{\bg}{\ensuremath{\bm{g}}\xspace}

\newcommand{\bh}{\ensuremath{\bm{h}}\xspace}

\newcommand{\bI}{\ensuremath{\mathbf{I}}\xspace}

\newcommand{\bp}{\ensuremath{\bm{p}}\xspace}

\newcommand{\bQ}{\ensuremath{\mathbf{Q}}\xspace}
\newcommand{\br}{\ensuremath{\bm{r}}\xspace}

\newcommand{\bT}{\ensuremath{\mathbf{T}}\xspace}
\newcommand{\bu}{\ensuremath{\bm{u}}\xspace}

\newcommand{\bw}{\ensuremath{\bm{w}}\xspace}

\newcommand{\bx}{\ensuremath{\bm{x}}\xspace}
\newcommand{\bX}{\ensuremath{\mathbf{X}}\xspace}
\newcommand{\by}{\ensuremath{\bm{y}}\xspace}

\newcommand{\bz}{\ensuremath{\bm{z}}\xspace}

\usepackage[final,bookmarks,bookmarksnumbered]{hyperref}
\hypersetup{%
    pdfborder = {0 0 0}
}

\begin{document}
\bstctlcite{IEEEexample:BSTcontrol}
\title{
	An Iterative Receiver for OFDM With Sparsity-Based Parametric Channel Estimation
}
\author{%
	Thomas L. Hansen, Peter B. Jørgensen, Mihai-Alin Badiu and Bernard H.
	Fleury%
	\thanks{T. L. Hansen, M.-A. Badiu and B. H. Fleury are with the Department
		of Electronic Systems at Aalborg University, Denmark. P. B. Jørgensen
		is with the Technical University of Denmark; this work was conducted
		while he was with Aalborg University.

	This work was supported by the Danish Council for Independent Research
	under grant ids DFF--4005-00549 and DFF--5054-00212 and by the European
	Commission in the framework of the FP7 Network of Excellence in Wireless
	COMmunications NEWCOM\# (Grant agreement no. 318306).  This work was also
	supported by the cooperative research project VIRTUOSO, funded by Intel
	Mobile Communications, Anite, Telenor, Aalborg University and Innovation
	Fund Denmark.

	\textcopyright 2018 IEEE. Personal use of this material is permitted.
	Permission from IEEE must be obtained for all other uses, in any current or
	future media, including reprinting/republishing this material for
	advertising or promotional purposes, creating new collective works, for
	resale or redistribution to servers or lists, or reuse of any copyrighted
	component of this work in other works.

	Digital Object Identifier 10.1109/TSP.2018.2868314
	}%
\vspace{-8mm}%
}
\markboth{IEEE Transactions on Signal Processing, Vol. 66, No. 20, Oct. 15,
2018}{Hansen \MakeLowercase{\textit{et al.}}:
An Iterative Receiver for OFDM With Sparsity-Based Parametric Channel Estimation}
\maketitle
\begin{abstract}
	In this work we design a receiver that iteratively passes soft information
	between the channel estimation and data decoding stages.
	The receiver incorporates sparsity-based parametric channel estimation.
	State-of-the-art sparsity-based iterative receivers simplify the channel
	estimation problem by restricting the multipath delays to a grid. Our
	receiver does not impose such a restriction. As a result it does not suffer
	from the leakage effect, which destroys sparsity. Communication at near
	capacity rates in high SNR requires a large modulation order. Due to the
	close proximity of modulation symbols in such systems, the grid-based
	approximation is of insufficient accuracy. We show numerically that a
	state-of-the-art iterative receiver with grid-based sparse channel
	estimation exhibits a bit-error-rate floor in the high SNR regime. On the
	contrary, our receiver performs very close to the perfect channel state
	information bound for all SNR values.
	We also demonstrate both
	theoretically and numerically that parametric channel estimation works well
	in dense channels, i.e., when the number of multipath components is large
	and each individual component cannot be resolved.
\end{abstract}


\section{Introduction}
Achieving high data-rate wireless communication with large spectral efficiency
requires the use of higher-order modulation formats, e.g. up to $256$-QAM in
3GPP LTE \cite{3gpp-36211}. Clearly using a high modulation order presuppose a
large signal-to-noise ratio (SNR), which will be supported by the envisioned
transition to small-cell operation. The availability of channel estimation
schemes that achieve high accuracy is crucial for receivers of systems with
large modulation order operating in the high-SNR regime.

To facilitate channel estimation, current systems embed pilot symbols into the
transmitted signal. In orthogonal frequency-division multiplexing (OFDM)
systems, a number of subcarriers are assigned to transmit pilot symbols.
The number of pilots is chosen to optimize throughput as a trade-off between
the amount of bandwidth and power allocated to pilot transmission and fidelity
of the channel estimate.

In this work we seek to improve upon this trade-off by designing a highly
accurate channel estimator while requiring a low pilot overhead. We propose a
unified receiver design that incorporates two main ideas: \textit{a)} an
iterative architecture and \textit{b)} sparsity-based parametric channel
estimation.

Our proposed receiver does not require any a-priori statistical information
about the wireless channel and is thus a particularly good candidate for
systems where no such information is available.

\subsection{Design of Iterative Receivers}
\begin{figure}[tbp]
	\centering\includegraphics[width=0.96\columnwidth]{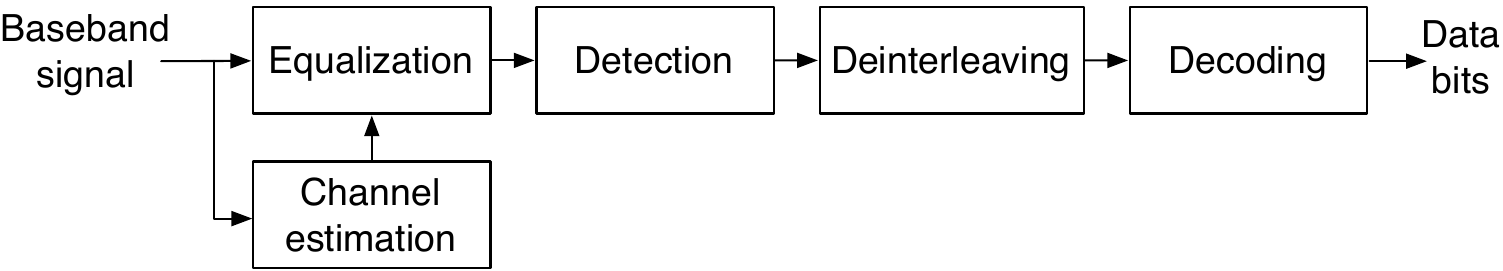}
	\caption{Flowchart of classical receiver design.}
	\label{fig:rx_flow}
\end{figure}

Classical receiver design employs a functional splitting of the process in the
receiver into independent subtasks, as illustrated in Fig. \ref{fig:rx_flow}.
Such a structure is suboptimal, since the information learned from the received
signal in any of the subtasks is only utilized in subsequent subtasks. To
remedy this sub-optimality feedback loops can be introduced between the
functional blocks in the receiver. This approach is known as the turbo
principle \cite{douillard-iterative, tuchler-turbo} due to its resemblance to
iterative decoding of turbo codes.




Application of the turbo principle has led to many iterative receiver designs,
e.g. \cite{douillard-iterative, tuchler-turbo, park-iterative}. Common to
these works is that each of the subtasks are designed independently using
traditional methods such as maximum likelihood (ML), maximum a-posteriori probability
(MAP) or minimum mean squared error (MMSE). The work
\cite{worthen-unified} introduced receiver design from the perspective of
inference in a factor graph. This allows for the receiver subtasks to be
designed \textit{jointly} with a certain objective in mind; a common example is
to seek the MAP estimate of the information bits.
Due to tractability and computational constraints, approximate inference
methods must be employed for iterative receiver design. Examples of
such methods are expectation propagation \cite{wu-expectation},
belief propagation (BP) with approximated messages \cite{zhu-message},
combined BP and mean-field (MF) \cite{riegler-merging, badiu-message},
relaxed BP \cite{schniter-belief} and generalized approximate
message-passing (GAMP) \cite{schniter-message}.

\subsection{Parametric Channel Estimation}
The impulse response of the compound channel (composed of the
transmitter RF front-end, the propagation channel and the receiver RF
front-end) is traditionally modelled as a sum of the form
\begin{align}
	g(\tau) = \sum_{l=1}^{L} \alpha_l c(\tau-\tau_l), \label{cir}
\end{align}
where $c(\tau)$ denotes the compound impulse response of the transmitter and
receiver RF front-ends. Here, $L$ is the number of so-called multipath
components. The $l$th multipath component is characterized by its coefficient or
weight $\alpha_l\in\bbc$ and its relative delay $\tau_l\in\bbr$. For short we
refer to $g(\tau)$ in \eqref{cir} as the channel impulse response (CIR). When
the number of multipath components $L$ is small relative to the number of OFDM
subcarriers, the model \eqref{cir} provides a parsimonious representation of
the compound channel and it is advantageous to perform channel estimation by
estimating the parameters of this model, i.e., estimating $L$, $\alpha_l$ and
$\tau_l$ for $l=1,\dots,L$. We refer to this approach as parametric channel
estimation.  It has been generally understood for many years
\cite{marple-resolution, candes-towards} that the delays can only be estimated
when the multipath components are well separated (see footnote
\ref{separated-footnote}). 

In our application context it is very restrictive to assume that the CIR takes
the form \eqref{cir} with $L$ small and all delays well separated. In
Sec.~\ref{sec:chan} we
demonstrate that even if the wireless channel exhibits a very large
number of (closely located) multipath components its CIR can still be
approximated by \eqref{cir} with $L$ small%
\footnote{
	Our analysis makes the usual assumption in OFDM of time-limited CIR, see
	\eqref{assumption}. Our results are therefore only directly applicable to
	scenarios where this assumption can reasonably be made, as is usually the
	case in radio communication.
}.
We refer to this approximation as the \textit{virtual} CIR.  We show that the
corresponding virtual channel frequency response (CFR) accurately approximates
the actual CFR within the system bandwidth. This means that we can use
\eqref{cir} with $L$ small as an estimation model even for channels of which
the multipath components cannot be resolved with the used system bandwidth.

Early works on parametric channel estimation address applications to
underwater communications \cite{feder-algorithms} and ultra-wideband (UWB)
communications \cite{lottici-channel, maravic-lowsampling}. Another classical
example is the rake receiver \cite{molisch-wideband}. All of these older works
assume that the number of (virtual) multipath components $L$ is known a priori or
use heuristics to estimate it.

A sparsity-based (or compressed sensing-based) approach can be used to allow
for inherent estimation of the number of (virtual) multipath components. Most
literature on sparsity-based channel estimation \cite{bajwa-sparsechannel,
paredes-uwbsparsechannel, berger-underwatersparsechannel,
taubock-sparsechannel, lovmand-sparsechannel, prasad-joint, prasad-joint-mimo}
employs a grid-based approximation of the CIR model \eqref{cir}, where the
multipath delays are confined to a discrete set of possible values. When a
baud-spaced grid%
\footnote{In the baud-spaced grid, the distance between adjacent
grid points is the reciprocal of the system bandwidth.}
is used, we refer to the samples of the CIR \eqref{cir} as channel taps.
The grid-based approximation results in a leakage effect%
\footnote{
	The compound wireless channel has a representation on the baud-spaced grid
	obtained by sampling $g(\tau)$. This representation is not sparse due to
	the presence of the RF front-end filter $c(\tau)$ in \eqref{cir} that
	introduces leakage.
	}
\cite{beek-channel, taubock-sparsechannel} and the vector of channel taps is
therefore only approximately sparse \cite{schniter-message, prasad-joint,
prasad-joint-mimo}. We demonstrate in our numerical investigation that the
grid-based approximation impairs the performance of receivers for OFDM systems
with large modulation order operating in the high-SNR regime. From a compressed
sensing point of view the effect of the grid-based approximation can be
understood as a basis mismatch \cite{chi-mismatchsensitivity}.

Recent works on off-grid compressed sensing have proposed methods that could in
principle be applied to sparsity-based channel estimation without resorting to
the grid approximation. These are based on atomic-norm minimization
\cite{candes-towards, bhaskar-atomic,pejoski-estimation}, finite rate of
innovation \cite{barbotin-fast} or Bayesian inference \cite{hansen-sparse,
badiu-valse, shutin-incremental, hu-compressed}. While all these methods show
good performance, the former two cannot easily be incorporated in an iterative
receiver. In this paper we show how sparsity-based parametric
channel estimation can be incorporated in an iterative receiver by
using approximate Bayesian inference. Our channel estimation scheme is
\textit{sparsity-based} in the sense that a sparsity-promoting prior model is used to
achieve inherent estimation of the number of (virtual) multipath components (the vector
\bz associated with \eqref{f_z} is sparse) and it is \textit{parametric} in the
sense that a parametric channel model is used to design the channel estimator.

\subsection{Prior Art}
Several prior works incorporate sparsity-based channel estimation in an
iterative
receiver. Prasad et. al. \cite{prasad-joint, prasad-joint-mimo} propose a joint
sparse channel estimation and \textit{detection} scheme for OFDM transmission.
Channel decoding is not considered in the joint processing and the EM algorithm
is used for channel inference. A baud-spaced grid is used.


Iterative receiver design for OFDM systems via GAMP and relaxed BP is proposed by
Schniter in \cite{schniter-belief,schniter-message}. The estimated multipath
delays are restricted to the baud-spaced grid. In the numerical evaluation of
\cite{schniter-belief} the CIRs fulfill this restriction, thus avoiding the
leakage effect at the expense of introducing an unrealistic channel model. In
\cite{schniter-message} a channel model generating continuous-valued delays is
assumed. It is shown that the channel taps follow a super-Gaussian density
that is modelled via a two-component Gaussian mixture. Due to the baud-spaced
grid the channel taps are correlated, which is mimicked with a hidden Markov
model. The resulting model has a large number of parameters to be estimated,
that causes systems with high-order modulation format to exhibit a
bit-error-rate (BER) floor when operating in the high-SNR regime (see
Sec~\ref{sec:numerical}).

The problem of parametric channel estimation based only on pilots or in the
contrived case when the data symbols are given is equivalent to that of line
spectral estimation \cite{lovmand-sparsechannel}. The work \cite{badiu-valse}
proposes a variational Bayesian approach to line spectral estimation. It is
shown that the Bernoulli-Gaussian prior \cite{kormylo-ml} is a powerful and
tractable sparsity-inducing model. Our sparsity-based parametric channel
estimator is inspired by \cite{badiu-valse} and uses the Bernoulli-Gaussian
prior model too. It differs from \cite{badiu-valse} in several aspects: \textit{a)}
at the data subcarriers the observations are modulated with the unknown data
symbols, \textit{b)} we impose that the estimate of the posterior probability
density function (pdf) of the
multipath coefficients factorizes and \textit{c)} to reduce computational
complexity we use a point estimate of the multipath delays.%
\footnote{By contrast, the scheme in \cite{badiu-valse} applied in our context
computes estimates of the posterior distribution of the delays.}

\subsection{Contributions}
The contributions of this paper are as follows:
\begin{enumerate}
	\item We propose a method to incorporate sparsity-based parametric channel
		estimation into an iterative receiver. Specifically we use the combined
		BP and MF (BP-MF) framework \cite{riegler-merging} to derive such an
		iterative receiver within a unified framework.
	\item We show (numerically) that iterative receivers for OFDM with high
		modulation order exhibit an error floor in the high-SNR regime when
		they employ state-of-the-art sparse channel estimation based on the
		baud-spaced grid approximation. Our iterative receiver design demonstrates
		how this error floor can be avoided.
	\item We demonstrate that parametric channel estimation, contrary to
		immediate intuition, can be applied to both specular and dense
		channels.
		In particular it is shown that the impulse response of any uncorrelated
		scattering channel can be approximated within the system bandwidth by
		a virtual CIR of the form \eqref{cir}. The number of
		components $L$ in the virtual CIR is equal to the effective rank of the
		channel covariance matrix.
		We demonstrate numerically that the
		effective rank of the channel covariance matrix is low for both a
		specular and a dense synthetic channel.
	\item Our algorithm development demonstrates how the BP-MF framework can be
		modified to provide approximate ML estimation of model parameters and
		how some latent variables can be estimated jointly to improve
		convergence speed. We expect that these approaches will prove useful in
		other applications of BP-MF.
\end{enumerate}
Our receiver only uses a few parameters (specifically the noise variance and
the two parameters of the Bernoulli-Gaussian prior model, sparsity level $\rho$
and multipath coefficient variance $\eta$) to describe the statistical
properties of the CIR and these are inherently estimated by
appropriately modifying BP-MF. This is in contrast to, for example, the linear
MMSE (LMMSE) channel estimators, which require a-priori specification of the
second-order statistics of the CFR \cite{edfors-ofdm,
park-iterative}, and the GAMP receiver \cite{schniter-belief}, which relies on
the second-order statistics of the channel taps and the transition
probabilities of the hidden Markov model.

The parametric channel estimation scheme that we propose requires the compound
frequency response of the RF front-ends to be known (at active
subcarriers). While that is not a wholly unrealistic assumption, it may prove
too restrictive in some practical situations. Since this frequency
response is stable over many OFDM symbols we expect that it can be estimated;
however that falls outside the scope of this work.

\subsection{Notation and Outline}
We denote column vectors as \ba and matrices as \bA. Transposition is denoted
as $(\cdot)\T$ and conjugate (Hermitian)
transposition as $(\cdot)\h$. The
scalar $a_i$ or $[\ba]_i$ gives the $i$th entry of vector \ba, while
$\ba_\cals$ gives a vector containing the entries in \ba at the indices in the
integer set \cals. The set difference operator $\cals\backslash\{i\}$ gives the
index set $\cals$ with index $i$ removed; we abuse notation slightly and write
$\cals\backslash i$ for short. The notation $[\bA]_{i,k}$ gives the $(i,k)$th
element of matrix \bA. We denote the vector \ba with the $i$th element removed
as $\ba_{\backslash i}$ and use a similar notation for matrices with columns
and/or rows removed (e.g. $[\bA]_{i, \backslash k}$ for the $i$th row with
$k$th entry removed).  The notation $\diag(\ba)$ denotes a matrix with the
entries of \ba on the diagonal and zeros elsewhere. The indicator function
$\mathbb{1}_{[\cdot]}$ gives $1$ when the condition in the brackets is
fulfilled and $0$ otherwise. The notation $a\proptoe b$ denotes $\exp(a)\propto
\exp(b)$, which implies $a = b + \const$ The multivariate complex normal
probability density function (pdf) is defined as
\begin{align*}
	\CN(\bx;\bmu,\bSigma) &\triangleq \pi^{-\dim(\bx)} |\bSigma|^{-1}
	\exp\!\left(-(\bx-\bmu)\h\bSigma\ii(\bx-\bmu)\right).
\end{align*}
The notation $\unif(x;0,T)$ gives the continuous uniform pdf on the interval
$[0,T]$ and $\mathrm{Bern}(x;\rho)$ gives the Bernoulli probability mass
function (pmf) for $x\in\{0,1\}$ with probability of success $\rho$.
We use $\ast$ to denote convolution and $\delta(\cdot)$ and $\delta[\cdot]$ to
denote the Dirac and Kronecker delta, respectively.

The paper is structured as follows: In Section \ref{sec:model} we specify the
observation model. In Section \ref{sec:inference} our approach to approximate
Bayesian inference is discussed. The inference algorithm is derived in detail
in Section \ref{sec:naive}. Section \ref{sec:numerical} presents the numerical
evaluation. Conclusions are given in Section \ref{sec:conclusions}.

\section{Modelling}\label{sec:model}

%

We consider data transmission using a single-input single-output OFDM system.
Since we do not exploit any structure between consecutive OFDM symbols, we
model the sequence of transmitted OFDM symbols to be independent and
identically distributed (i.i.d.).  The OFDM system transmits $P$ pilot
subcarriers and $D$ data subcarriers, such that the total number of subcarriers
per symbol is $N=P+D$. The sets $\calp$ and \cald give the indices of the pilot
and data subcarriers, respectively. It follows that $\mathcal{D}\cup\mathcal{P}
= \{1,\dots,N\}$ and $\mathcal{D}\cap\mathcal{P} = \emptyset$.

\subsection{OFDM System}
\label{sec:ofdmsystem}
The $K$ (equi-probable) information bits to be transmitted are stacked in vector
$\bu\in\{0,1\}^K$.
These bits are coded by a rate-$R$ encoder and interleaved to get the length-$K/R$
vector $\bc=\calc(\bu)$. The interleaving and coding function
$\calc:\{0,1\}^K\rightarrow\{0,1\}^{K/R}$ can represent any interleaver and
coder, e.g. a turbo \cite{berrou-turbo}, low-density parity check (LDPC)
\cite{gallager-ldpc} or convolutional code. We split \bc into subvectors
$\bc^{(i)}\in\{0,1\}^Q$, ${i\in\cald}$, such that $\bc^{(i)}$ contains the $Q$ bits
that are mapped to the $i$th subcarrier. The complex symbols
$x_i=\calm(\bc^{(i)})$, ${i\in\cald}$, are obtained via the $2^Q$-ary mapping
$\calm:\{0,1\}^Q\rightarrow\bba_|D|\subset\bbc$, where $\bba_|D|$ is the data symbol
alphabet. The pilots are selected in the pilot symbol alphabet
$\bba_|P|\subset\bbc$. In OFDM, $\bba_|D|$ is typically a $2^Q$-ary quadrature
amplitude modulation (QAM)
alphabet and $\bba_|P|$ a quadrature phase shift keying (QPSK) alphabet. The
pilot and data symbols are stacked in vector $\bx$. Vector $\bx_\cald$
contains the data symbols and $\bx_\calp$ contains the pilot symbols.

The transmitter and receiver are assumed to operate with perfect time
synchronization. We also assume that the local oscillators in the transmitter
and receiver are perfectly synchronized and that these oscillators are ideal
(i.e., no phase noise, etc.) We consider a baseband signal model and assume
ideal conversion to and from the carrier-frequency passband signal.
The RF front-ends are modelled as linear time-invariant filters with compound
impulse response $c(\tau)=c_|TX|(\tau) \ast c_|RX|(\tau)$. The wireless channel
is also assumed linear and time-invariant for the duration of an OFDM symbol.
The impulse response of the propagation channel during transmission of the
current OFDM symbol is denoted $h(\tau)$ and the (compound) CIR is then
\begin{align}
	g(\tau) = c(\tau) \ast h(\tau).
	\label{g_tau}
\end{align}
We make the usual assumption of time-limited CIR:
\begin{align}
	g(\tau) = 0\quad \textrm{ for } \quad\tau\notin[0,T_|CP|],
	\label{assumption}
\end{align}
where $T_|CP|$ is the cyclic prefix duration. In practice this assumption needs
only to be fulfilled relative to the noise level.%
\footnote{Specifically the signal
contribution in \eqref{obs} arising from the tail of the compound CIR outside $[0,T_|CP|]$
should be neglectable compared to noise.}

By the assumption in \eqref{assumption} the OFDM system operates without
inter-symbol interference, so we can consider transmission of a single OFDM
symbol. The OFDM transmitter is modelled as a baseband processor followed by an
RF front-end that applies the filter $c_|TX|(\tau)$. The baseband processor
emits
\begin{align}
	s(t) =
	\begin{cases}
		\sum_{n=1}^N x_n \exp(j2\pi\Delta_f n t)
		& t\in[-T_|CP|, T_|sym|] \\
		0 & \text{otherwise},
	\end{cases}
\end{align}
where $\Delta_f$ gives the subcarrier spacing and $T_|sym|=\Delta_f\ii$ is
the OFDM symbol length. The OFDM receiver is modelled as an RF front-end that
applies the filter $c_|RX|(\tau)$ followed by a baseband processor that
samples the signal.
The signal at the output of the receiver RF front-end is
\begin{align}
	r(t) = g(\tau) \ast s(t) + w(t),
\end{align}
where $w(t)$ is low-pass filtered white Gaussian noise.
The receiver baseband processor samples $r(t)$, removes the cyclic prefix and calculates the
discrete Fourier transform to obtain the observed vector \by. The assumption in
\eqref{assumption} ensures that orthogonality of the subcarriers is preserved.
It can be shown \cite{luise-blind} that
\begin{align}
	\by = \bX\bg + \bw,
	\label{obs}
\end{align}
where $\bX=\diag(\bx)$. The Gaussian noise vector \bw is assumed%
\footnote{This assumption is fulfilled when the receive RF
front-end has constant frequency response within the system bandwidth.} white
with component variance $\beta$. The vector \bg contains samples of the
compound CFR at the subcarrier frequencies and its entries are
\begin{align}
	g_n = \int_{0}^{T_|sym|} \hspace{-.5em}g(\tau) \exp(-j2\pi\Delta_fn \tau)
	\,\dd\tau,
	\quad n=1,\dots,N.
	\label{g_n}
\end{align}
Inserting \eqref{g_tau} into \eqref{g_n} and by
the convolution theorem we can obtain (see \cite{luise-blind} for details)
\begin{align}
	\bg = \bC\bh,
\end{align}
where $\bC=\diag(\bc)$. The vectors $\bc$ and $\bh$ contain samples of the
Fourier transform of $c(\tau)$ and $h(\tau)$, respectively. These vectors are
obtained analogously to \eqref{g_n}.

\subsection{Parametric Channel Model}
\label{sec:chan}
We now consider a model for the propagation channel
$h(\tau)$. A classical model is the uncorrelated-scattering (US)
channel \cite{bello-characterization}, in which $h(\tau)$ is modelled as a
stochastic process with autocorrelation
\begin{align}
	\E{h(\tau) h^*(\tau')} = \rho(\tau) \delta(\tau-\tau').
\end{align}
The function $\rho(\tau)$ is the power-delay profile (PDP). We further
assume that the process $h(\tau)$ is zero-mean.
The vector \bh is then also zero-mean. Denote the covariance matrix of \bh as
$\bSigma=\E{\bh\bh\h}$.
Using the US assumption it can be shown that the frequency-response vector \bh contains
samples of a wide-sense-stationary random process and that \bSigma is a
Toeplitz matrix.

Denote the rank of the $N\times N$ matrix \bSigma as $L$.
Then the Carathe\'odory parameterization of a Toeplitz matrix
\cite{stoica-spectral, caratheodory-zusammenhang}
states that there exist vectors $\btau\in[0,\Delta_f\ii)^L$ and
$\bgamma\in[0,\infty)^L$ such that
\begin{align}
	\bSigma = \bPsi(\btau)\bGamma\bPsi(\btau),
	\label{bSigma}
\end{align}
where $\bGamma=\diag(\bgamma)$ and the matrix $\bPsi(\btau)\in\bbc^{N\times L}$
has $(n, l)$th entry $\exp(-j2\pi\Delta_fn \tau_l)$, $n=1,\dots,N$,
$l=1,\dots,L$. Note that the parameterization is unique if and only if $L<N$.
From \eqref{bSigma} it is clear that $\bh$ lies in the column space of
$\bPsi(\btau)$ and that it can be represented as
$\bh = \bPsi(\btau)\balpha$ for some $\balpha\in\bbc^L$. It then follows that
\begin{align}
	\bg = \bC\bPsi(\btau)\balpha.
	\label{g}
\end{align}
The parametric channel estimator that we employ is obtained by estimating \btau and
\balpha in the above parametric model of \bg.
It is recognized that \bh is a superposition of complex sinusoids.
Thus, given \bX and \bC, the estimation of $L$, \balpha and \btau reduces to an
instance of line spectral estimation.

The reuse of notation between \eqref{cir} and \eqref{g} is not accidental. If
the CIR is assumed to take the parametric from \eqref{cir} and this
CIR is Fourier transformed to obtain the CFR, we get exactly the expression
\eqref{g}. Parametric channel estimators are in fact usually motivated by
assuming that the CIR has the from \eqref{cir}.
But in the above we showed that the parametric model \eqref{g} can be obtained
from the US assumption, i.e., without explicitly imposing a model of the form
\eqref{cir}. This means that parametric channel estimation can be used for all
US channels. The pair $(\tau_l, \alpha_l)$ denotes the delay and complex
coefficient of a virtual multipath component.
The $L$ virtual multipath components described by $(\btau, \balpha)$ can be
inserted into \eqref{cir} to obtain a virtual CIR. The above shows that if the
covariance matrix $\bSigma$ indeed has rank $L$ the corresponding virtual CFR
coincides with the CFR of the actual channel within the system bandwidth.

In this work we make the simplifying assumption that the filters in the RF
front-ends have constant frequency response within the system bandwidth.
This assumption means that $\bC=\bI$ (any constant scaling can be integrated
into \balpha).
This assumption is reasonable because typical OFDM systems employ a number of
unused virtual (or guard) subcarriers in the roll-off region of the RF
front-end filters \cite{sari-transmission}.

\begin{remark}
	The assumption $\bC=\bI$ does not mean that there are no filters at RF
	front-ends. It just means that these filters have unit frequency response
	within the system bandwidth. To be precise, the wireless channel is
	``observed'' by the receiver as described by \eqref{obs} and \eqref{g}.
	It is clear that the wireless channel is observed only within a
	band-limited interval of length $N\Delta_f$.
\end{remark}
\begin{remark}
	The assumption $\bC=\bI$ can be relaxed to the assumption that the
	frequency response of the RF front-ends is arbitrary but known or
	estimated by the receiver (i.e., the matrix \bC is known or estimated)
	\cite{barbu-sparse}. Since \bC is fixed across many OFDM symbols we expect
	that it can be estimated with high accuracy by the receiver. An
	investigation of such an approach is outside the scope of this paper. If
	\bC is known the derivation in Sec.~\ref{sec:naive} can be
	straightforwardly extended to include \bC. We
	use $\bC=\bI$ to keep the notation simple.
\end{remark}

\subsection{Specular and Dense Channels}
\label{sec:specular-dense}
The rank of the channel covariance matrix $L$ describes the channel's number of
degrees of freedom. The smaller this number, the fewer parameters are needed in
\eqref{g} to describe the channel and the higher channel estimation accuracy
can be achieved. We are thus particularly interested in the case where \bSigma
is low-rank.

In this paper we classify channels into the two categories of specular and dense
channels. For specular channels the CIR truly has the form \eqref{cir} with $L$ much smaller than
$N$ and the delays in $\btau$ are well separated.%
\footnote{``Well separated'' is here meant relative to the reciprocal of the
system bandwidth $1/(N\Delta_f)$.\label{separated-footnote}}
In such channels the delays \btau and coefficients \balpha can directly be
estimated as indicated by \cite{candes-towards}. It is easy to show that the
channel covariance matrix of a specular channel does indeed take the form
\eqref{bSigma} and that it has low rank. Empirical evidence suggests that the
wireless channel in some propagation environments is specular to a large
extent. In practice, specular channels are composed of a small number of
dominant multipath components and a remaining part with power below the noise
floor. Examples include the ultra-wideband channels that are considered for 5G
wireless communications \cite{cramer-evaluation, molisch-ultrawideband} and
underwater acoustic channels \cite{stojanovic-underwater}. See also
\cite{berger-sparsechannel, bajwa-sparsechannel} and references therein.

It is, however, broadly accepted that wireless channels are not always specular
\cite{turin-communication, schniter-message, bajwa-sparsechannel}. In the
general case they are composed of a very large number of multipath
components that do not adhere to a minimum separation condition. That is caused
by diffuse scattering and by rich scattering environments.  We refer to such
channels as dense. In dense channels it is not possible to estimate the delay
and coefficient of each multipath component in \eqref{cir}.  The use of the
Carathe\'odory parameterization shows that it is, however, still possible to
estimate a set of virtual multipath components that approximate the actual CFR
within the system bandwidth. As discussed above the parametric approach works
better when \bSigma has low-rank or, in other words, when the virtual CIR has
only few components.  Using a representation based on discrete prolate
spheroidal sequences \cite{slepian-dpss} it can be shown that the assumption
\eqref{assumption} implies that \bh effectively lies in a subspace with
dimension approximately given by $\ceil{T_|CP|N\Delta_f}$. This value then also
gives an upper bound on the effective rank%
\footnote{By the effective rank we mean a rank that ignores very small
eigenvalues.}
of \bSigma. OFDM systems are practically always designed
such that $T_|CP|\Delta_f\ll1$ and so \bSigma has low effective rank.  In many
cases the effective rank is even lower than $\ceil{T_|CP|N\Delta_f}$. We
demonstrate below that this is the case for a standardized and widely used
model of a dense channel.

\begin{figure}[t]
	\setlength\figureheight{32mm}
	\setlength\figurewidth{70mm}
	\pgfplotsset{local_axis_style/.style={
		legend pos = south east,
		grid = both,
		xmin=0, xmax=175,
		xtick distance=25
	}} 
	\centering
%
\begin{tikzpicture}

\begin{axis}[%
width=0.951\figurewidth,
height=\figureheight,
at={(0\figurewidth,0\figureheight)},
scale only axis,
xmin=0,
xmax=180,
xlabel={Effective rank of $\bSigma$},
ymin=0,
ymax=1,
ylabel={Cumulative probability},
axis background/.style={fill=white},
legend style={legend cell align=left,align=left,draw=white!15!black},
global_axis_style, local_axis_style
]
\addplot[const plot,color=blue,solid] plot table[row sep=crcr] {%
9	0\\
9	0.003\\
10	0.00900000000000001\\
11	0.02\\
12	0.036\\
13	0.048\\
14	0.0619999999999999\\
15	0.094\\
16	0.133\\
17	0.179\\
18	0.246\\
19	0.327\\
20	0.447\\
21	0.586\\
22	0.72\\
23	0.89\\
24	1\\
};
\addlegendentry{Scenario A};

\addplot[const plot,color=red,dashed] plot table[row sep=crcr] {%
38	0\\
38	0.001\\
42	0.003\\
43	0.004\\
44	0.005\\
47	0.00600000000000001\\
48	0.00700000000000001\\
49	0.00900000000000001\\
50	0.011\\
51	0.015\\
52	0.018\\
53	0.02\\
54	0.025\\
55	0.027\\
56	0.032\\
57	0.037\\
58	0.038\\
59	0.041\\
60	0.045\\
62	0.046\\
63	0.048\\
64	0.051\\
65	0.0600000000000001\\
66	0.0650000000000001\\
67	0.0680000000000001\\
68	0.0740000000000001\\
69	0.0790000000000001\\
70	0.0900000000000001\\
71	0.0930000000000001\\
72	0.103\\
73	0.115\\
74	0.122\\
75	0.128\\
76	0.148\\
77	0.156\\
78	0.164\\
79	0.174\\
80	0.184\\
81	0.195\\
82	0.213\\
83	0.228\\
84	0.236\\
85	0.252\\
86	0.263\\
87	0.278\\
88	0.29\\
89	0.297\\
90	0.311\\
91	0.318\\
92	0.328\\
93	0.346\\
94	0.365\\
95	0.38\\
96	0.394\\
97	0.413\\
98	0.425\\
99	0.447\\
100	0.467\\
101	0.489\\
102	0.511\\
103	0.54\\
104	0.561\\
105	0.58\\
106	0.597\\
107	0.611\\
108	0.628\\
109	0.642\\
110	0.654\\
111	0.669\\
112	0.68\\
113	0.705\\
114	0.723\\
115	0.738\\
116	0.757\\
117	0.771\\
118	0.79\\
119	0.804\\
120	0.816\\
121	0.829\\
122	0.844\\
123	0.861\\
124	0.878\\
125	0.892\\
126	0.902\\
127	0.907\\
128	0.912\\
129	0.924\\
130	0.933\\
131	0.94\\
132	0.948\\
133	0.955\\
134	0.963\\
136	0.966\\
137	0.969\\
138	0.974\\
139	0.978\\
140	0.979\\
141	0.984\\
142	0.988\\
143	0.989\\
145	0.99\\
146	0.991\\
147	0.992\\
148	0.994\\
149	0.995\\
151	0.996\\
155	0.997\\
156	0.998\\
160	0.999\\
161	1\\
};
\addlegendentry{Scenario B};

\end{axis}
\end{tikzpicture}%
%
	\vspace{-2mm}%
	\caption{Empirical CDFs of effective rank of $\bSigma$ when a threshold is
	applied to the eigenvalues. The plot is obtained by estimating the CDF from
	$1,000$ realizations in each of the scenarios described in
	Sec.~\ref{sec:numerical}.}
	\label{fig:rank}
\end{figure}
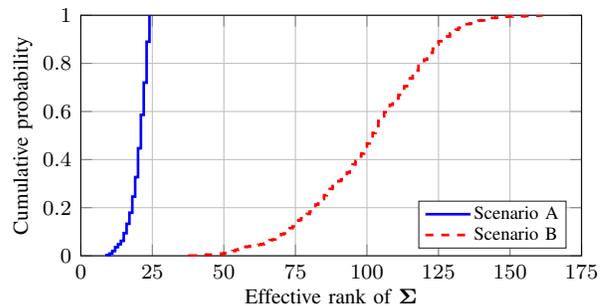

To investigate the effective rank of the channel covariance matrix we conduct a
numerical experiment. We here give a quick overview of this experiment and
refer to \cite[Chap. 3]{hansen-thesis} for more details. The experiment is
based on the two propagation scenarios described in Sec.~\ref{sec:numerical}.
The channel model of Scenario A is specular to a large extent while the channel
model of Scenario B is dense. The channel models randomly generate a set of
multipath delays and component powers that describe the local behaviour of the
channel. The channel covariance matrix is obtained by inserting these delays
and powers into \eqref{bSigma}. The obtained \bSigma has many eigenvalues that
are small but still non-zero (this is not caused by limited numerical
precision). We therefore define a method to neglect small eigenvalues that do
not represent significant power. For that purpose the eigenvalues are
normalized such that their average value is 1. The normalized eigenvalues below
$10^{-4}$ are then set to zero. This means that whenever the SNR is
significantly below $40\db$, the removed power is neglectable in comparison to
the noise power. Even in perfect conditions wireless communication systems
practically always operate significantly below 40 dB SNR. For that reason we
find this to be a conservative approach to thresholding the eigenvalues. Taking
the resulting number of non-zero eigenvalues gives the effective rank.
Fig.~\ref{fig:rank} depicts empirical cumulative distribution functions (CDFs)
of the effective rank in Scenario A and B. From Table~\ref{tab:parameters} we have
$\ceil{T_|CP|N\Delta_f}=134$ in Scenario A and $\ceil{T_|CP|N\Delta_f}=205$. It
is seen that the effective rank of \bSigma is generally much smaller than
$\ceil{T_|CP|N\Delta_f}$.
In Scenario B the effective rank is also much smaller
than the number of multipath components in the channel, indicating that a
virtual CIR of the form \eqref{cir} with $L$ small exists even for dense
channels.
Due to the use of the effective rank (and not the true rank) of the channel
covariance matrix, the virtual CFR approximates the true CFR. The approximation
is only accurate within the system bandwidth.

In summary it can be concluded that parametric channel estimation can be
applied to both specular and dense channels. That is indeed confirmed in the
numerical investigation reported in Sec.~\ref{sec:numerical}.

\begin{figure*}[t]
	\centering
	\includegraphics[scale=0.75]{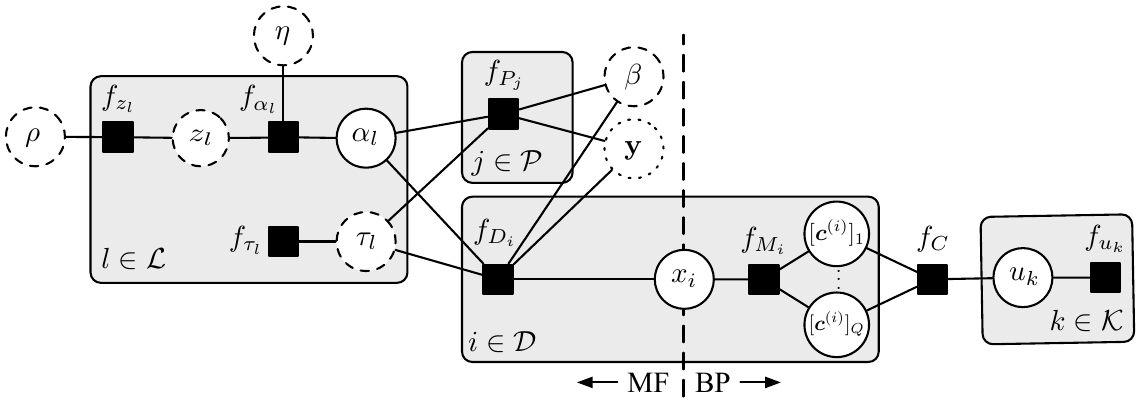}
	\caption{Factor graph representation of the probabilistic model describing
	the complete OFDM system and channel model. The shaded areas indicate
	multiple copies of the nodes, as specified by the index sets. The vector of
	observations \by is included with a dotted line because it is known at the
	time of inference. Variables of which a point estimate is obtained (as opposed to a
	full variational estimate of the posterior pdf) are represented by circles
	with dashed line. The vertical dashed line shows the separation between the BP and
	MF subgraphs.}
	\label{fig:fg}
\end{figure*}

\subsection{Probabilistic Model of the OFDM System}
\label{sec:factorgraph}
We are now ready to present a probabilistic model that describes the complete
OFDM system. The model expresses the joint probability of all variables in the
system as a product of factors. This factorization of the joint probability is
represented as the factor graph depicted in Fig. \ref{fig:fg}. The factor graph
representation is central in the formulation of the receiver algorithm. In the following
we introduce the variables and factors in the factor graph, moving from right
to left.%
\footnote{We abuse terminology and associate variables and factors with their
respective nodes in the factor graph.}

The interleaving, coding and modulation of the data bits are described in Sec.
\ref{sec:ofdmsystem}. The subgraph characterizing the system implementing these
tasks involves the factors
\begin{align*}
	f_{u_k}(u_k) &\triangleq p(u_k)
		= 0.5\, \mathbb{1}_{[u_k\in\{0,1\}]},
		& &{k\in\calk}, \\
	f_C(\bc, \bu) &\triangleq p(\bc|\bu)
		= \mathbb{1}_{[\bc = \calc(\bu)]}, \\
	f_{M_i}(x_i, \bc^{(i)}) &\triangleq p(x_i | \bc^{(i)})
	= \mathbb{1}_{[x_i = \calm(\bc^{(i)})]},
		& &{i\in\cald},
\end{align*}
where $\calk=\{1,\dots,K\}$ is the index set of the information bits. The
factor $f_C(\bc,\bu)$ describes the interleaving and channel coding processes. By
``zooming in'' this factor can be expanded to a subgraph involving auxiliary
variables and factors that describe the structure of the channel code and
interleaver.

The subgraph characterizing the observation process described by \eqref{obs}
and \eqref{g} involves the following factors for pilot- and data subcarriers,
respectively:
\begin{align*}
	f_{P_j}(\balpha, \btau, \beta)
		&\triangleq p(y_j | \balpha, \btau; \beta) \\
		&= \CN(y_j ; x_j[\bPsi(\btau)\balpha]_j, \beta),
		& &{j\in\calp}, \\
	f_{D_i}(x_i, \balpha, \btau, \beta)
		&\triangleq p(y_i | x_i, \balpha, \btau; \beta) \\
		&= \CN(y_i ; x_i[\bPsi(\btau)\balpha]_i, \beta),
		& &{i\in\cald}.
\end{align*}

The $l$th virtual multipath components is modelled through the variables
$\alpha_l$, $\tau_l$ and $z_l$. To ease the terminology we drop the attribute
``virtual'' in the following.
To model the fact that there are only a multipath components, a
Bernoulli-Gaussian prior is used. This prior assigns large probability to the
event $\alpha_l=0$. The model contains $L_|max|$ multipath components of which
only a subset is activated, i.e. has $\alpha_l\ne0$. The number $L_|max|$ is an
upper bound on the number of multipath components that can be estimated.%
\footnote{In our implementation we select $L_|max|=\ceil{T_|CP|N\Delta_f}+1$,
which is the maximum number of degrees of freedom under the assumption
\eqref{assumption} \cite{bajwa-sparsechannel}. It roughly corresponds to the
number of baud-spaced (spacing $1/(N\Delta_f)$) components on the interval
$[0,T_|CP|)$.}
This allows us to derive an algorithm that inherently estimates the number of
multipath components. Each component is assigned an activation variable
$z_l\in\{0,1\}$, which is $1$ when said multipath component is active and $0$
otherwise. The sequence $\{z_1,\ldots,z_{L_|max|}\}$ is modelled i.i.d. where each
$z_l$ is assigned a Bernoulli prior with activation probability $\rho$:
\begin{align}
	f_{z_l}(z_l, \rho) &\triangleq p(z_l; \rho)
		= \mathrm{Bern}(z_l; \rho),
		& &{l\in\call},
		\label{f_z}
\end{align}
where we have defined the set of multipath component indices
$\call=\{1,\dots,L_|max|\}$. 
The prior density of the multipath coefficient $\alpha_l$ is conditioned on $z_l$, such that $z_l=0$
implies $\alpha_l=0$ and $z_l=1$ gives a Gaussian density with variance $\eta$:
\begin{align*}
	f_{\alpha_l}(\alpha_l, z_l, \eta) &\triangleq p(\alpha_l|z_l; \eta) \\
	&= (1-z_l)\delta(\alpha_l)
		+ z_l\CN(\alpha_l ; 0, \eta),
		& &{l\in\call}.
\end{align*}
When performing inference in this model, the estimated number of active multipath
components is $\hat L \triangleq \norm{\hat\balpha}_0$, where $\hat\balpha$ is a
vector containing the estimates of $\alpha_l$ for all $l\in\call$.

We finally need to impose a prior model on the multipath delays $\tau_l$,
$l\in\call$. The only prior information available is through the assumption
\eqref{assumption} that implies that for all $l\in\call$ we have
$0\le\tau_l\le T_|CP|$. To express this an i.i.d. uniform prior is used:
\begin{align*}
	f_{\tau_l}(\tau_l) &\triangleq p(\tau_l)
		= \unif(\tau_l ; 0, T_|CP|),
		& &{l\in\call}.
\end{align*}

\section{Inference Method}\label{sec:inference}

The BER optimal receiver (assuming $\rho$, $\eta$ and $\beta$ known) computes
the MAP estimate
\begin{align}
	\hat u_k = \argmax_{u_k\in\{0,1\}}\;
		p(u_k | \by; \rho, \eta, \beta), \quad{k\in\calk}.
\end{align}
The pdf $p(u_k|\by; \rho, \eta, \beta) \propto p(u_k, \by; \rho, \eta, \beta)$
can ideally be found by marginalizing all variables but $u_k$ in the joint pdf
\begin{multline*}
	p(\by, \bz, \balpha, \btau, \bx_\cald, \bc, \bu; \rho, \eta, \beta) =
	p(\by | \bx_\cald, \balpha, \btau; \beta) \\
	\prod_{l\in\call}p(\alpha_l|z_l; \eta) p(z_l; \rho) p(\tau_l)
	\prod_{i\in\cald}p(x_i|\bc^{(i)}) p(\bc|\bu)
	\prod_{k\in\calk}p(u_{k}).
\end{multline*}
Calculating the marginals of $u_k$, $k\in\calk$, is intractable and we resort
to approximate Bayesian inference.

\subsection{Combined Belief Propagation and Mean-Field}
Our inference method is based on the merged belief propagation and mean-field
(BP-MF) framework of \cite{riegler-merging}. In this framework a so-called
belief function is found for each variable in the factor graph. The belief
function is an approximation of the marginal posterior pdf or pmf of that variable. We
abuse notation and let $q(a)$ denote the belief of variable $a$. When the set
of belief functions has been calculated, the MAP estimate of the $k$th data
bit is found as the mode of $q(u_k)$.

For tractability we obtain a point estimate of the variables \bz and \btau.
This is achieved as proposed in \cite{riegler-merging} by restricting their
beliefs to be Kronecker and Dirac delta functions, i.e.,
$q(z_l)=\delta[z_l-\hat z_l]$ and $q(\tau_l)=\delta(\tau_l-\hat\tau_l)$ for all
$l\in\call$.


At the heart of BP-MF lies the so-called region-based free energy approximation
(RBFE) \cite{yedidia-constructing}. The RBFE is obtained by splitting the
factor graph into a MF and a BP subgraph, as indicated in Fig. \ref{fig:fg}. The RBFE
is a function of%
\footnote{The RBFE is also a functional of the beliefs corresponding to the
factors in the BP subgraph. BP-MF enforces consistency between the variable
beliefs and these factor beliefs. Since the latter are not relevant to the
derivation of the receiver, we omit them.}
the point estimates $\hat\bz,\hat\btau$ and the belief functions $q(\alpha_l),
q(x_i), q([\bc^{(i)}]_m)$ and $q(u_k)$ for indices
$l\in\call,i\in\cald,k\in\calk$ and $m=1,\dots,Q$. It is also a function of the
model parameter estimates $(\hat\rho,\hat\eta,\hat\beta)$, as justified
below. The expression of the RBFE is given in Appendix \ref{app:rbfe}.
BP-MF seeks to minimize the RBFE under a number of normalization and
consistency constraints. The messages of BP-MF are derived such that at
convergence they satisfy the Karush-Kuhn-Tucker conditions of the constrained
RBFE minimization, i.e., a (possibly local) minimum of the constrained problem
is found. See \cite{riegler-merging} for a more detailed discussion of BP-MF.

The understanding of BP-MF as RBFE minimization allows us to make a number of
adaptations to the message-passing scheme to improve convergence speed.
Further, we will see that this understanding is useful when analyzing
convergence of the algorithm.

\subsection{Model Parameter Estimation with BP-MF}
The BP-MF framework \cite{riegler-merging} does not directly provide a method
to estimate the unknown model parameters $(\rho,\eta,\beta)$. We propose to do
so by letting the RBFE be a function of these model parameters. The model
parameter estimates $(\hat\rho,\hat\eta,\hat\beta)$ are then obtained as the
minimizers of the RBFE.

To justify this method we first note that the model parameters are located in
the MF subgraph. Then we follow an approach similar to \cite{winn-vmp} to
obtain a lower bound on the marginal log-likelihood function:
\begin{align}
	\ln p(\by; \hat\rho,\hat\eta,\hat\beta) \ge
		- F_|BP-MF|
		+ \const,
		\label{evidence_bound}
\end{align}
where $F_|BP-MF|$ is the RBFE \eqref{rbfe} and the constant only depends on
beliefs of variables in the BP subgraph (including $q(x_i)$, for $i\in\cald$),
i.e., it does not depend on $(\hat\rho, \hat\eta, \hat\beta)$. It can then be
seen that the values of $(\hat\rho,\hat\eta,\hat\beta)$ minimizing $F_|BP-MF|$
maximize the lower bound on the likelihood function in \eqref{evidence_bound}.
These minimizers are thus approximate ML estimates.  We note that if the above
approach is applied in a pure MF context it simplifies to variational EM
estimation with all other variables treated as latent variables
\cite{riegler-merging, beal-phd}.

\subsection{Relation to Prior Art}
To relate our receiver algorithm to current methods we note that the decoding
of many popular channel codes can be described as an instance of BP
\cite{pearl-bp} in a factor graph \cite{kschischang-factor, mceliece-turbo,
kschischang-iterative}. For example, BP decoding of a convolutional code
leads to the BCJR algorithm \cite{bahl-bcjr}. We see in Fig. \ref{fig:fg} that
the merged BP-MF algorithm employs BP in the subgraph that represents the
channel code, i.e., standard techniques are used for decoding.

Similarly, there are examples in the literature of MF inference where the
underlying factor graph resembles the MF subgraph of our receiver. The work
\cite{badiu-valse} uses a Bernoulli-Gaussian prior model similar to that in our
work, while \cite{hansen-sparse}, \cite{shutin-incremental} use a
gamma-Gaussian prior typical of sparse Bayesian learning.

The strength of the BP-MF framework is now clear: It allows us to merge
existing methods for channel decoding and sparsity-based estimation using a
unifying design method (namely that of RBFE minimization).

\section{Parametric BP-MF Receiver}
\label{sec:naive}
To minimize the RBFE, we apply the BP-MF algorithm given by Eq. (21)--(22) in
\cite{riegler-merging} on the factor graph of Fig.  \ref{fig:fg}. In the
following we use the notation $\left<\cdot\right>_{a}$ to denote expectation
with respect to the belief density $q(a)$. We follow the convention of
\cite{riegler-merging} in naming the messages. In \cite{badiu-message} a
similar BP-MF receiver is derived, which does not exploit channel sparsity.

\subsection{Message Passing for Channel Estimation}
\subsubsection{Update of Coefficient Belief}
We start by finding belief updates in the MF subgraph.
To find the update of
$q(\alpha_l)$, $l\in\call$, we calculate the messages passed to the node $\alpha_l$:
\begin{align*}
	m_{f_{\alpha_l}\rightarrow\alpha_l}^\textrm{MF}(\alpha_l)
	&\propto
		\begin{cases}
			\ee{-\hat\eta\ii |\alpha_l|^2}		& \text{if } \hat z_l = 1, \\
			\delta(\alpha_l)					& \text{if } \hat z_l = 0
		\end{cases}\\
	m_{f_{D_i}\rightarrow\alpha_l}^\textrm{MF}(\alpha_l)
	&\propto
		\ee{-\hat\beta\ii
		\left<|y_i - x_i[\bPsi(\hat\btau)\balpha]_i|^2\right>_{x_i,\balpha_{\backslash l}}} \\
	m_{f_{P_j}\rightarrow\alpha_l}^\textrm{MF}(\alpha_l)
	&\propto
		\ee{-\hat\beta\ii
			\left<|y_j - x_j[\bPsi(\hat\btau)\balpha]_j|^2\right>_{\balpha_{\backslash l}}},
\end{align*}
which holds for all $l\in\call$, $i\in\cald$ and $j\in\calp$.
Taking the product of all messages going into the node $\alpha_l$ gives its belief
\begin{align}
		q(\alpha_l) &=
		\begin{cases}
	\CN(\alpha_l ; \hat\mu_l, \hat\sigma^2_l)	& \text{if } \hat z_l = 1 \\
			\delta(\alpha_l)					& \text{if } \hat z_l = 0
		\end{cases}
	\label{naive-alpha}
\end{align}
with the active component mean and variance
\begin{align}
	\hat\mu_l &= \hat\sigma^2_l q_l
		\label{naive-mu} \\
	\hat\sigma^2_l &= \left( s_l + \hat\eta\ii \right)\ii.
		\label{naive-sigma}
\end{align}
Here we have introduced
\begin{align}
	s_l &= \hat\beta\ii \bpsi\h(\hat\tau_l)
		\left<\bX\h\bX\right>_{\bx_{\cald}} \bpsi(\hat\tau_l)
		\label{naive-sl} \\
	q_l &= \hat\beta\ii \bpsi\h(\hat\tau_l) \br
		\label{naive-ql} \\
	\br &= \left<\bX\right>\h_{\bx_{\cald}} \by
		- \left<\bX\h\bX\right>_{\bx_{\cald}}
		\bPsi(\hat\btau_{\hat\cala\backslash l}) \hat\bmu_{\hat\cala\backslash l}, \label{r}
\end{align}
where $\bpsi(\tau_l)$ is defined as the $l$th column of $\bPsi(\btau)$.
Note that the belief of inactive components ($\hat z_l=0$) becomes a point mass
at $\alpha_l=0$, thus eliminating the influence of that component in the
product $\bX\bPsi(\hat\btau)\balpha$. We have defined the set of currently active
components as $\hat\cala\triangleq\{l:\hat z_l=1\}$ and the vectors
$\hat\bmu=[\hat\mu_1, \dots, \hat\mu_{L_|max|}]\T$, $\hat\bsigma^2=[\hat\sigma_1^2,
\dots, \hat\sigma_{L_|max|}^2]\T$.

\subsubsection{Joint Update of Delay and Coefficient Belief}
\label{subsec:naive-delays}
We now turn our attention to the estimation of the multipath delays $\tau_l$,
$l\in\call$.
To improve the convergence speed of the algorithm, we find the update of
$\hat\tau_l$ by minimizing the RBFE \textit{jointly} with respect to the
beliefs $q(\alpha_l)$ and $\hat\tau_l$.
Due to the selected prior $p(\tau_l)$, the following expressions are valid for
$\hat\tau_l\in[0,T_|CP|]$. We are only concerned with active components, i.e.,
$l\in\hat\cala$ and thus $\hat z_l=1$.
Writing only the terms of the RBFE \eqref{rbfe} that depend on $q(\alpha_l)$ and
$\hat\tau_l$, we get
\begin{align}
	F_|BP-MF|(q(\alpha_l),\hat\tau_l) \proptoe
		\int q(\alpha_l)
		\ln \frac{q(\alpha_l)}{Q(\alpha_l,\hat\tau_l)} \,\dd \alpha_l
		\label{tau-rbfe}
\end{align}
with
{\small
\begin{align}
	Q(\alpha_l, \hat\tau_l) &=
		p(\alpha_l|\hat z_l;\hat\eta)p(\hat\tau_l)
		\exp\!\left(\left< \ln p(\by|\bx_\cald, \balpha, \hat\btau;\hat\beta)
			\right>_{\bx_\cald,\balpha_{\backslash l}}\right)
		\nonumber \\
	&\propto \CN(\alpha_l ; \hat\mu_l, \hat\sigma^2_l)
		\exp\!\left( \frac{|q_l|^2}{s_l + \hat\eta\ii} \right),
	\label{Q_taualpha}
\end{align}
}%
where $\hat\sigma^2_l$, $\hat\mu_l$, $s_l$ and $q_l$ are given by
\eqref{naive-mu} - \eqref{naive-ql} and thus implicitly are functions of
$\hat\tau_l$.
We need to minimize \eqref{tau-rbfe} under the normalization constraint $\int
q(\alpha_l) \,\dd \alpha_l = 1$. To do so, define
\begin{align}
	g_{\tau_l}(\hat\tau_l) &\triangleq \qquad \max_
	{\mathclap{\tilde q(\alpha_l):\int \tilde q(\alpha_l)\,\dd\alpha_l=1}}
	\;\; - F_|BP-MF|\left(\tilde q(\alpha_l),
		\hat\tau_l\right) \label{f_tau} \\
		&\proptoe \ln\int Q(\alpha_l,\hat\tau_l)\,\dd\alpha_l \label{f_tau_2} \\
	&\proptoe \frac{\hat\beta^{-2}}{s_l + \hat\eta\ii} |\bpsi\h(\hat\tau_l) \br|^2.
	\label{f_tau_final}
\end{align}
The result in \eqref{f_tau_2} is easily obtained by noting that
\eqref{tau-rbfe} can be rewritten as
\begin{align*}
	F_|BP-MF| \proptoe
		\mathrm{KL}\left[q(\alpha_l) \Big|\Big|
			\frac{Q(\alpha_l,\hat\tau_l)}{\int
		Q(\tilde\alpha_l,\hat\tau_l)\,\dd\tilde\alpha_l}\right]
		- \ln\int Q(\tilde\alpha_l,\hat\tau_l)\,\dd\tilde\alpha_l,
	\end{align*}
where
$\mathrm{KL}[\cdot||\cdot]$ is the Kullback-Leibler divergence. The coefficient
belief is selected as the maximizer of \eqref{f_tau},
i.e., $q(\alpha_l)=Q(\alpha_l,\hat\tau_l) / \int
Q(\tilde\alpha_l,\hat\tau_l)\,\dd\tilde\alpha_l$, which is easily shown to
coincide with the result in \eqref{naive-alpha}.

Since $s_l$ is constant with respect to $\hat\tau_l$, we find the delay update as
\begin{align}
	\hat\tau_l
		&= \argmax_{\tilde\tau_l\in[0,T_|CP|]}\;
			g_{\tau_l}(\tilde\tau_l)
		= \argmax_{\tilde\tau_l\in[0,T_|CP|]}\;
			|\bpsi\h(\tilde\tau_l) \br|^2.
		\label{tau-update}
\end{align}

We recognize the objective function in \eqref{tau-update} as the continuous
periodogram of the residual vector \br.
While it is possible to find the maximizer of the periodogram, doing so has
high computational cost. In our iterative algorithm, we instead find an update
of $\hat\tau_l$ that cannot increase the objective in \eqref{tau-update}.
Denote the updated delay estimate as $\hat\tau_l^{[t]}$ and the previous delay
estimate as $\hat\tau_l^{[t-1]}$. Our scheme now reads:
\begin{enumerate}
	\item Find initial step
			$\Delta
				= \frac{g_\tau'(\hat\tau_l^{[t-1]})}{|g_\tau''(\hat\tau_l^{[t-1]})|}$.
		\item If $g_\tau(\hat\tau_l^{[t-1]}+\Delta)\ge g_\tau(\hat\tau_l^{[t-1]})$, set
		$\hat\tau_l^{[t]} = \hat\tau_l^{[t-1]} + \Delta$ and
		terminate.
		Otherwise set $\Delta = \frac{\Delta}{2}$ and repeat step 2.
\end{enumerate}
Functions $g_\tau'(\tau_l)$ and $g_\tau''(\tau_l)$ are the first and second derivatives
of \eqref{f_tau_final}. The scheme gives the Newton update of
$\hat\tau_l$ if this value increases the objective function and otherwise
resorts to a gradient ascent with a backtracking line search.
We have the following lemma, that we will use in the convergence analysis:


\begin{lemma}
	\label{lem:tauupdate}
	The procedure listed in Steps 1-2 above followed by an update of $q(\alpha_l)$
	does not increasing the RBFE.
\end{lemma}
\begin{IEEEproof}
	First, note that the updated $\hat\tau_l$, does not decrease
	$g_{\tau_l}(\hat\tau_l)$. It then follows that by selecting the maximizer
	of \eqref{f_tau}, the RBFE is non-increasing.
\end{IEEEproof}

\subsubsection{Joint Update of Activation Variable and Coefficient Belief}
We now turn our focus on the update of the activation variable $\hat z_l$. It is
again desirable to perform a joint update of $\hat z_l$ and $q(\alpha_l)$.
We proceed in a similar way as we did to compute the updates of the multipath delays. The terms in
the RBFE \eqref{rbfe}, which depend on $q(\alpha_l)$ and $\hat z_l$, are denoted as
$F_|BP-MF|(q(\alpha_l), \hat z_l)$. We then define
\begin{align}
	g_{z_l}(\hat z_l) &\triangleq \qquad \max_
	{\mathclap{\tilde q(\alpha_l):\int \tilde q(\alpha_l)\,\dd\alpha_l=1}}
	\;\; - F_|BP-MF|(\tilde q(\alpha_l),
		\hat z_l) \label{g_z} \\
	&\proptoe
		\begin{cases}
			\frac{|\hat\mu_l|^2}{\hat\sigma_l^2} + \ln
			\frac{\hat\sigma_l^2}{\hat\eta} +  \ln\hat\rho
					& \text{if } \hat z_l = 1, \\
			\ln(1-\hat\rho)
					& \text{if } \hat z_l = 0.
		\end{cases}
\end{align}
This result is easily obtained by following steps analogous to
\eqref{tau-rbfe} -- \eqref{f_tau_final}.
The activation variable solves the decision problem
$\hat z_l = \max_{\tilde z_l\in\{0,1\}} g_{z_l}(\tilde z_l)$.
Writing the ``activation criterion'' $g_{z_l}(1) > g_{z_l}(0)$ we get
\begin{align}
	\frac{|\hat\mu_l|^2}{\hat\sigma_l^2}
		> \ln\frac{\hat\eta}{\hat\sigma_l^2} + \ln \frac{1-\hat\rho}{\hat\rho}.
		\label{criterion}
\end{align}
If the above criterion is true we set $\hat z_l=1$; otherwise we set $\hat
z_l=0$. The corresponding update of $q(\alpha_l)$ is the maximizer of
\eqref{g_z}, which remains as in \eqref{naive-alpha}. The
criterion in \eqref{criterion} is the same as that obtained in
\cite{badiu-valse}.

\subsubsection{Update of Channel Parameter Estimates}
The channel parameters $(\rho, \eta, \beta)$ are estimated as the values that
minimize the RBFE. Writing only the terms of the RBFE \eqref{rbfe} that depend
on the channel parameters we have
\begin{flalign*}
	F_|BP-MF|(\hat\rho,\hat\eta,\hat\beta) & &
\end{flalign*}
\vspace{-6mm}
\begin{multline*}
	\proptoe - \left< \ln \prod_{l\in\call}p(\hat z_l;\hat\rho)p(\alpha_l|\hat
	z_l;\hat\eta)
	p(\by|\balpha,\hat\btau,\bx_\cald;\hat\beta)\right>_{\bx_\cald, \balpha}
	\hspace{-2em} \hfill
\end{multline*}
\vspace{-6mm}
\begin{multline}
	\proptoe \norm{\hat\bz}_0\ln\hat\rho + (L_|max|-\norm{\hat\bz}_0)\ln(1-\hat\rho)
		- N\ln\hat\beta - \hat\beta\ii u \\
	- \norm{\hat\bz}_0\ln\hat\eta
		- \hat\eta\ii\sum_{\{l:\hat z_l = 1\}} (|\hat\mu_l|^2+\hat\sigma_l^2),
		\label{modelparameters}
\end{multline}
where
\begin{flalign*}
	u &\triangleq \left< \norm{\by-\bX\bPsi(\hat\btau)\balpha}^2
		\right>_{\bx_\cald,\balpha} & \\
	&= \norm{\by}^2 
	+ \hat\bmu_{\hat\cala}\h\bPsi\h(\hat\btau_{\hat\cala})
	\left<\bX\h\bX\right>_{\bx_\cald}
\bPsi(\hat\btau_{\hat\cala}) \hat\bmu_{\hat\cala}
\end{flalign*}
\vspace{-6mm}
\begin{gather*}
	+ \sum_{l\in\hat\cala}\hat\sigma_l^2 \bpsi\h(\hat\tau_l)
		\left<\bX\h\bX\right>_{\bx_{\cald}} \bpsi(\hat\tau_l)
\end{gather*}
\vspace{-6mm}
\begin{flalign}
	&& - 2\re\!\left\{ \by\h \left<\bX\right>_{\bx_\cald} \bPsi(\hat\btau_{\hat\cala})
	\hat\bmu_{\hat\cala}\right\}.
\end{flalign}
It is readily seen that $F_|BP-MF|(\hat\rho,\hat\eta,\hat\beta)$ can be minimized independently
with respect to each of the parameters. By taking derivatives and equating to
zero we find the global minima (the second derivatives are all positive):
\begin{align}
	\hat\rho &= \frac{\norm{\hat\bz}_0}{L_|max|} \label{rho-update} \\
	\hat\eta &= \frac{\sum_{\{l:\hat z_l=1\}}
		(|\hat\mu_l|^2 + \hat\sigma_l^2)}{\norm{\hat\bz}_0} \label{eta-update} \\
	\hat\beta &= \frac{u}{N}. \label{beta-update}
\end{align}

\subsubsection{Iterating all Coefficient Beliefs Ad-Infinitum}
In \cite{shutin-incremental} it is demonstrated that 
iterating the updates of some variables \textit{ad-infinitum} is a powerful
technique for increasing the convergence speed of MF algorithms.
We apply that idea to the beliefs of the multipath coefficients.

Since $q(\alpha_l)=\delta(\alpha_l)$ for all $l\in\call\backslash\hat\cala$,
the following discussion is only concerned with the beliefs of active
components, i.e. for $l\in\hat\cala$.  First note that the variance \eqref{naive-sigma} of an active
multipath coefficient $\hat\sigma_l^2$ does not depend on the beliefs of the
remaining coefficients $q(\alpha_k)$, $k\ne l$. The mean \eqref{naive-mu} of
the $l$th coefficient, on the other hand, depends on the remaining mean values
as
\begin{multline*}
	\hat\mu_l = \underbrace{\hat\sigma_l^2}_{[\bQ]_{l,l}\ii}
		\bigg(
			\underbrace{\hat\beta\ii \bpsi\h(\hat\tau_l)
				\langle\bX\rangle\h_{\bx_{\cald}} \by}_{p_l} \\
		- \sum_{k\in\hat\cala\backslash l}
			\underbrace{ \hat\beta\ii \bpsi\h(\hat\tau_l)
			\langle\bX\h\bX\rangle_{\bx_{\cald}}
			\bpsi(\hat\tau_k) }_{[\bQ]_{l,k}}
		\hat\mu_k \bigg)
\end{multline*}
for all $l\in\hat\cala$. The matrix $\bQ$ is of size $|\hat\cala|\times|\hat\cala|$ and we
have abused notation in using $l,k$ as indices into this matrix, because $1\le
l,k\le L_|max|$, even though $|\hat\cala|\le L_|max|$. The above equation is recognized as the
Gauss-Seidel \cite{golub-matrix} iteration for solving the system of linear
equations
\begin{align}
	\bQ \hat\bmu_{\hat\cala} &= \bp \label{naive-bmu-update}
\end{align}
with
\begin{align*}
	\bp &= \hat\beta\ii \bPsi\h(\hat\btau_{\hat\cala})
		\left<\bX\right>_{\bx_{\cald}}\h \by \\
		\bQ &= \hat\beta\ii \bPsi\h(\hat\btau_{\hat\cala})
		\left<\bX\h\bX\right>_{\bx_{\cald}}
		\bPsi(\hat\btau_{\hat\cala})
		+ \hat\eta\ii\bI.
\end{align*}
It follows that the updates of $\hat\mu_l$, for all $l\in\hat\cala$, converge to the
solution $\hat\bmu_{\hat\cala}$ found by solving \eqref{naive-bmu-update}.

We note that in the hypothetical special case where the beliefs of \bX are
point estimates (or equivalently known)
$\by=\bX\bPsi(\hat\btau_{\hat\cala})\balpha_{\hat\cala} + \bw$ is a linear observation model with
Gaussian noise. In this case, the estimator $\hat\bmu_{\hat\cala}=\bQ\ii\bp$ reduces
to the LMMSE estimator of $\balpha_{\hat\cala}$ in the linear observation model under
the Bayesian model dictated by the current beliefs of the remaining variables.
The estimator $\hat\bmu_{\hat\cala}=\bQ\ii\bp$ is, however, not the LMMSE estimator
of $\balpha_{\hat\cala}$ when the uncertainty of the estimate of \bX is considered.

\subsection{Message-Passing for Decoding}
In the previous subsections we derived the \textit{belief functions} $q(\cdot)$
of the variables whose factor neighbours are in the MF subgraph only.
To perform inference in the BP subgraph, i.e., detection, demapping, decoding
and deinterleaving, we need to calculate the \textit{messages} that
are passed along its edges.

We begin with the messages $n_{x_i\rightarrow f_{M_i}}(x_i)$, $i\in\cald$, which constitute
the interface from the continuous-valued channel estimator to the
discrete-valued decoder. They are given as
\begin{multline}
	n_{x_i\rightarrow f_{M_i}}(x_i) = m_{f_{D_i}\rightarrow
		x_i}^\textrm{MF}(x_i) \\
		\propto
		\CN\!\left( x_i ;
			\frac{ y_i \left< g_i \right>_{\balpha,\btau}^*  }
				{\left<\big|g_i\big|^2\right>_{\balpha,\btau}},
		\frac{\hat\beta}{\left<\big|g_i\big|^2\right>_{\balpha,\btau}} \right),
		\label{message-x-fM}
\end{multline}
where $g_i \triangleq [\bPsi(\btau)\balpha]_i$ is the CFR sampled at subcarrier
$i$. Its mean and second moment are
\begin{align*}
	\left< g_i \right>_{\balpha,\btau} &= [\bPsi(\hat\btau) \hat\bmu]_i \\
		\left<\big|g_i\big|^2\right>_{\balpha,\btau}
		&= \big[ \bPsi(\hat\btau) (\hat\bmu\hat\bmu\h + \diag(\hat\bsigma^2))
			\bPsi\h(\hat\btau) \big]_{i,i}.
\end{align*}
Note that even though the above expression has the form of a Gaussian, the
messages are probability mass functions obtained by evaluating the above
Gaussian at the points of the symbol alphabet $\bba_|D|$ followed by appropriate
normalization.

The mean in \eqref{message-x-fM} can be interpreted as the output of an LMMSE
equalizer. Consider the observation model $y_i=g_ix_i+w_i$ where
$p(w_i)=\CN(w_i;0,\hat\beta)$ and $g_i = [\bPsi(\hat\btau)\balpha]_i$. Let
$q(\alpha_l)$ be the density of $\alpha_l$ and impose a prior $p(x_i) =
\CN(x_i; 0,\sigma_{x_i}^2)$ on $x_i$. The LMMSE estimator of $x_i$ is now
\begin{align*}
	\hat x_i^{\textrm{LMMSE}} &=
	\frac{y_i \left< g_i \right>_{\balpha,\btau}^*}
		{\left<\big|g_i\big|^2\right>_{\balpha,\btau}
		+ \hat\beta\sigma_{x_i}^{-2}}.
\end{align*}
By letting $\sigma_{x_i}^2\rightarrow\infty$ to express that we have no prior
information on $x_i$, we recover the mean in \eqref{message-x-fM}. Note that a
similar analogy does not exist for the variance in \eqref{message-x-fM}.

All remaining messages passed in the BP subgraph are functions of discrete
variables (i.e., coded or information bits). These messages are calculated with
the sum-product algorithm, see e.g.  \cite{kschischang-factor, mceliece-turbo}.
Due to space constraints, we do not give the details here.

When BP messages have been passed in the BP subgraph, the beliefs of the data
symbols $x_i$, $i\in\cald$, are calculated from
\begin{align}
	q(x_i) &\propto
		m_{f_{D_i}\rightarrow x_i}^\textrm{MF}(x_i)
		m_{f_{M_i}\rightarrow x_i}^\textrm{BP}(x_i).
		\label{x-update}
\end{align}
Since $q(x_i)$ is a probability mass function, we can use straightforward
evaluation of finite sums to obtain $\left<\bX\right>_{\bx_\cald}$ and
$\left<\bX\h\bX\right>_{\bx_\cald}$, which are used in the belief updates in
the MF subgraph.

\subsection{An Incremental Algorithm}
Algorithm \ref{alg:naive} combines the derived belief update expressions into
an iterative receiver with sparsity-based parametric channel estimation. The
algorithm is split into two parts: channel estimation (lines
\ref{line:mf-begin} - \ref{line:mf-end}) and decoding (line
\ref{line:bp-part}). The outer loop alternates between these two steps until
the information bit estimates have not changed in $10$ iterations or a maximum
of $50$ iterations is reached.

\begin{figure}[t]
\removelatexerror
\begin{algorithm}[H]
	\small
	\caption{Parametric BP-MF receiver.}
	\label{alg:naive}
	\SetAlgoLined
	\DontPrintSemicolon
	\KwIn{Observations \by, pilot indices $\calp$ and pilot symbols $\bx_{\calp}$.}
	\KwOut{Belief functions of data bits $\{q(u_k)\}_{k\in\calk}$.}
	\KwNotes{Define the set of components as $\call=\{1,\ldots,L_|max|\}$ and the set
		of active components as
		$\hat\cala\triangleq\{l\in\call:\hat z_l=1\}$.}
	$\tilde\btau \gets$ Vector with values from equispaced grid on
		$[-\frac{1/2}{N\Delta_f}, T_|CP|]$. \hspace{-.1em}\;
	Initialize channel parameter estimates $(\hat\rho, \hat\eta, \hat\beta)$. \;
	$\hat\bz, \hat\btau, \hat\bmu, \hat\bsigma^2 \gets $ Zero vectors of length $N$. \;
	\While{Outer stopping criterion not met}{
		\While{Inner stopping criterion not met}{
			\label{line:mf-begin}
			$\hat\bmu_{\hat\cala},\hat\bsigma^2_{\hat\cala} \gets$ Updates from 
				\eqref{naive-bmu-update} and \eqref{naive-sigma}. \;
				\label{line:naive-add-bmu}
			\textit{Activate an inactive component:} \;
				\label{line:add-start}
			\If{the inactive set $\call\backslash\hat\cala$ is non-empty}{
				$l \gets $ Any index from the inactive set $\call\backslash\hat\cala$. \;
				$\hat z_l \gets 1$. \;
				$\hat\tau_l \gets$ Value from \eqref{tau-update} calculated
					on the grid $\tilde\btau$. \;
					\label{line:gridsearch}
				$\hat\bmu_{\hat\cala},\hat\bsigma^2_{\hat\cala} \gets$ Updates from 
					\eqref{naive-bmu-update} and \eqref{naive-sigma}. \;
				$\hat\tau_l \gets$ Update via the scheme in Sec.
					\ref{subsec:naive-delays}. \;
					\label{line:tau-update-add}
				$\hat\mu_l, \hat\sigma^2_l \gets$ Updates from \eqref{naive-mu}
					and \eqref{naive-sigma}. \;
				\If{activation criterion \eqref{criterion} is false}{
					$\hat z_l \gets 0$. \;
					Reset $\hat\bmu_{\hat\cala}$ to the value calculated in line
							\ref{line:naive-add-bmu}. \;
				}
			}

			\textit{Update all components currently included in model:} \;
			\label{line:refine-start}
			\For{$l\in\hat\cala$}{
				\label{line:refine-loop}
				$\hat\tau_l \gets$ Update via the scheme in Sec.
					\ref{subsec:naive-delays}. \;
					\label{line:tau-update-refine}
				$\hat\mu_l, \hat\sigma^2_l \gets$ Updates from \eqref{naive-mu}
					and \eqref{naive-sigma}. \;
				\If{activation criterion \eqref{criterion} is false}{
					$\hat z_l \gets 0$. \;
				}
			}

			$\hat\bmu_{\hat\cala},\hat\bsigma^2_{\hat\cala} \gets$ Updates from 
				\eqref{naive-bmu-update} and \eqref{naive-sigma}. \;

			$\hat\rho,\hat\eta,\hat\beta \gets$ Updates from
			\eqref{rho-update}, \eqref{eta-update} and \eqref{beta-update}. \;
		} 
			\label{line:mf-end}
		Update the messages $m_{f_{D_i}\rightarrow x_i}^\textrm{MF}(x_i)$ from \eqref{message-x-fM}. \;
		Iterate message-passing in the BP subgraph. \;
			\label{line:bp-part}
		Update the beliefs $q(x_i)$ from \eqref{x-update}. \;
	} 
\end{algorithm}
\end{figure}

The scheduling of the channel estimation is inspired by \cite{hansen-sparse}.
The basic idea is to construct a representation of the CFR in \eqref{g} by
sequential refinement of the estimated multipath components. One
component is determined by the parameters $(z_l, \alpha_l, \tau_l)$ for
a particular index $l$.  All multipath components are initialized in the
inactivated state, i.e., $\hat\bz$ is the zero vector.

The channel estimation procedure alternates between two stages: In the
activation stage (at line \ref{line:add-start}) one of the inactive components
is activated and its multipath delay and coefficient are calculated. The
activation criterion \eqref{criterion} determines if the component should stay
activated.
In the second stage (starting at line \ref{line:refine-start}), all active
components are sequentially refined. Again, the criterion \eqref{criterion}
determines if a component should be deactivated. The channel estimation
procedure thus
iteratively adds, updates and possibly removes components until the stopping
criterion is fulfilled.
The multipath delays are tracked via the scheme in Sec.
\ref{subsec:naive-delays} in a way that resembles the operation of a rake
receiver \cite{molisch-wideband}. The approach presented here differs from that
implemented in a rake receiver in that it provides an integral criterion for
inclusion or exclusion of components (rake ``fingers'') via \eqref{criterion}.
The multipath delay of the newly activated component is found via a
maximization over the grid $\tilde\btau$. The grid should have a sufficiently
fine resolution, such that the initial estimate of the delay is close to the
maximizer in \eqref{tau-update}. We choose the distance between points in
the grid as $(N\Delta_f)\ii/8$.
As inner stopping criterion we use $|1/\hat\beta^{[t]} - 1/\hat\beta^{[t-1]}| <
10^{-3}/\hat\beta^{[t-1]}$, where $t$ is the inner iteration number. The number
of inner iterations is limited to 50.

During the first outer iteration the decoder has not been used yet and symbol
beliefs $q(x_i)$ of the data subcarriers (indices $i\in\cald$) are not
available. During the first iteration the channel estimator therefore only
uses the pilot subcarriers (indices $j\in\calp$). To avoid any identifiability
issue regarding the multipath delays (see Sec.~\ref{sec:pilotspacing}) during the
pilot-only iteration, the multipath delays estimated in this iteration are
restricted to the interval $[0,1/(\Delta_f\Delta_P))$, where $\Delta_P$ is the
pilot spacing.%
\footnote{We define the pilot spacing as $\Delta_P=D+1$, where $D$ is the number of
data subcarriers between any two neighboring pilot subcarriers.}

The active component prior variance is initialized to $\hat\eta=1$ and the
activation probability is initialized to $\hat\rho=0.5$. We initialize the
noise variance to $\hat\beta=\norm{\by}^2/N\cdot10^{-15/10}$ (i.e., assuming
approximately $15\db$ SNR). The activation probability and noise variance is
kept fixed during the first $3$ outer iterations, because these can only be
accurately estimated when a reliable estimate of the channel is available.

%

\begin{table*}[t]
	\centering
	\vspace{7pt}
	\begin{tabularx}{\textwidth}{Xrr}
		\toprule
		\textbf{Parameter} & \textbf{Scenario A} & \textbf{Scenario B} \\
		\midrule
		Channel model							& ITU-R M.2135 UMa NLOS
												\cite{itu-m2135}
												& IEEE 802.15.a Outdoor NLOS
												\cite{molisch-comprehensive} \\
		Number of subcarriers ($N$)				& $1024$
												& $1024$ \\
		Modulation format of data subcarriers	& 256-QAM
												& 256-QAM \\
		Convolutional code polynomial			& $(561,753)_8$
												& $(561,753)_8$ \\
		Subcarrier spacing ($\Delta_f$)			& $25\;$kHz
												& $250\;$kHz \\
		Cyclic prefix duration ($T_|CP|$)		& $5200\;$ns
												& $800\;$ns \\
		Number of equispaced pilots				& $172$
												& $256$ \\
		Pilot spacing ($\Delta_P$) (implied by the above)	& $6$
												& $4$ \\
		\bottomrule
	\end{tabularx}
	\vspace{1.5mm}
	\caption{Simulation parameters.}
	\label{tab:parameters}
\end{table*}

\subsection{Convergence Analysis and Computational Complexity}
%

We now wish to analyze the convergence properties of Algorithm \ref{alg:naive}.
First recognize that the algorithm alternates between updates in the MF and BP
subgraphs of Fig. \ref{fig:fg}. To analyze convergence, we discuss under which
conditions each of these sets of updates are guaranteed not to increase the
RBFE. If all updates give a non-increasing RBFE it can be concluded that the
algorithm converges, since the RBFE is bounded below.


We first discuss the updates in the MF subgraph, i.e., of belief functions
$q(\alpha_l)$ ($l\in\call$) and point estimates ($\hat\bz,\hat\btau,\hat\rho,
\hat\eta, \hat\beta$). During these updates the messages $m^\text{BP}_{f_{M_i}\rightarrow
x_i}(x_i)$ are kept fixed. The joint update of $\hat\tau_l$ and $q(\alpha_l)$
gives a non-increasing RBFE as per Lemma \ref{lem:tauupdate}. A similar
conclusion can be drawn regarding the joint update of $\hat z_l$ and
$q(\alpha_l)$. The individual update of $q(\alpha_l)$ is found via the method
of Lagrange multipliers applied to the RBFE with normalization constraint $\int
q(\alpha_l)\,\dd\alpha_l=1$. The second-order functional derivative of the RBFE
$\frac{\delta^2 F_\text{BP-MF}}{\delta q^2(\alpha_l)}=\frac{1}{q(\alpha_l)}$ is
a positive semi-definite function; it follows that the RBFE is convex in this
argument.  It can be concluded that the update of $q(\alpha_l)$ is the global
minimizer of the RBFE and the objective is thus non-increasing. A similar
conclusion can be drawn regarding the update of the channel parameters, cf. Eq.
\eqref{modelparameters}. All updates in the MF subgraph thus give
non-increasing RBFE.

We now analyze the convergence in the BP subgraph, i.e., the updates of belief
functions $q(x_i),q([\bc^{(i)}]_q)$ and $q(u_k)$. Considering the belief
functions of variables in the MF subgraph as fixed and ignoring scaling and
constant terms, the RBFE is equal to the Bethe free energy corresponding to the
factorization (see \cite[Appendix E]{riegler-merging})
\begin{align*}
	\prod_{i\in\cald}
		m^\text{MF}_{f_{D_i}\rightarrow x_i}(x_i) p(x_i|\bc^{(i)})
		p(\bc|\bu) \prod_{k\in\calk} p(u_k).
\end{align*}
Further, all messages in the BP subgraph are equal to the messages obtained
from BP applied to the above factorization.
This means that we can analyze the behaviour of message-passing in the BP
subgraph, by analyzing BP applied to the above factorization. If the factor
graph does not contain any cycles it can be shown that BP globally minimizes the
Bethe free energy \cite{yedidia-constructing,riegler-merging} (which in this
case is equal to the Gibss free energy) and convergence of the complete BP-MF
receiver algorithm is guaranteed.
Recall that the factor $f_C(\bc,\bu)=p(\bc|\bu)$ describes the channel code and may be
replaced by a number of auxiliary variables and factors. The specific structure
of the BP factor graph is thus determined by the channel code. In the special
case of convolutional coding with binary or quadrature phase-shift
keying (BPSK or QPSK) modulation, the BP graph does indeed become a tree-graph
and convergence of Alg. \ref{alg:naive} is guaranteed. If the modulation order
is higher than QPSK, loops occur between $f_{M_i}$ and $f_C$ and convergence
can thus not be guaranteed.

For other common channel codes, such as Turbo and LDPC codes, the subgraph
represented by $f_C$ contains loops. However, BP has empirically been shown to
converge for decoding of many channel codes and it is a well known practice to
use BP even though convergence cannot be guaranteed theoretically, see e.g.
\cite{kschischang-factor, mceliece-turbo, kschischang-iterative, bahl-bcjr}.
When BP does converge it has been shown to be to a (local) minimum of the Bethe
free energy \cite{heskes-stable}, which further explains why we do indeed
obtain convergence of Alg. \ref{alg:naive} in our numerical investigations.
Conditions exist under which BP is guaranteed to converge in loopy
graphs, e.g. \cite{mooij-sufficient,su-convergence}. These are, however, not
applicable to our situation.

We now turn our attention to the computational complexity of the channel
estimator, i.e., the loop starting at line \ref{line:mf-begin}. The most
demanding part of the channel estimation in terms of computational complexity
is the calculation of $\hat\bmu_{\hat\cala}$ via \eqref{naive-bmu-update}.
We show in Appendix \ref{app:bmu} that (under a conjecture) this update can be
calculated in time $\calo\!\left( \min(\hat L^2N, \hat LN\sqrt{N}) \right)$,
where $\hat L$ is the number of components currently included in the model.

The grid search in line \ref{line:gridsearch} is recognized as the maximization
of the periodogram, which can be calculated via a fast Fourier transform in
time $\calo\!\left( N\log N \right)$ when the grid is assumed to be of size
$\calo(N)$.

The loop starting at line \ref{line:refine-loop} necessitates the calculation
of \br in \eqref{r}. Direct computation has complexity $\calo(\hat
LN)$ for each of the $\hat L$ iterations in the loop. By updating $\br$ with
each change to $\hat\bmu$, the direct evaluation can be avoided and the
complexity of each iteration in the loop becomes $\calo(N)$, which is the same
as that of all other operations inside the loop. The overall complexity of the
loop is thus $\calo(\hat LN)$.

With these remarks, we see that the overall complexity per iteration of the
channel estimator is $\calo\!\left( \min(\hat L^2N, \hat LN\sqrt{N}) \right)$.

\section{Numerical Evaluation} \label{sec:numerical}

%
%
%
%
%

\begin{figure*}[t]
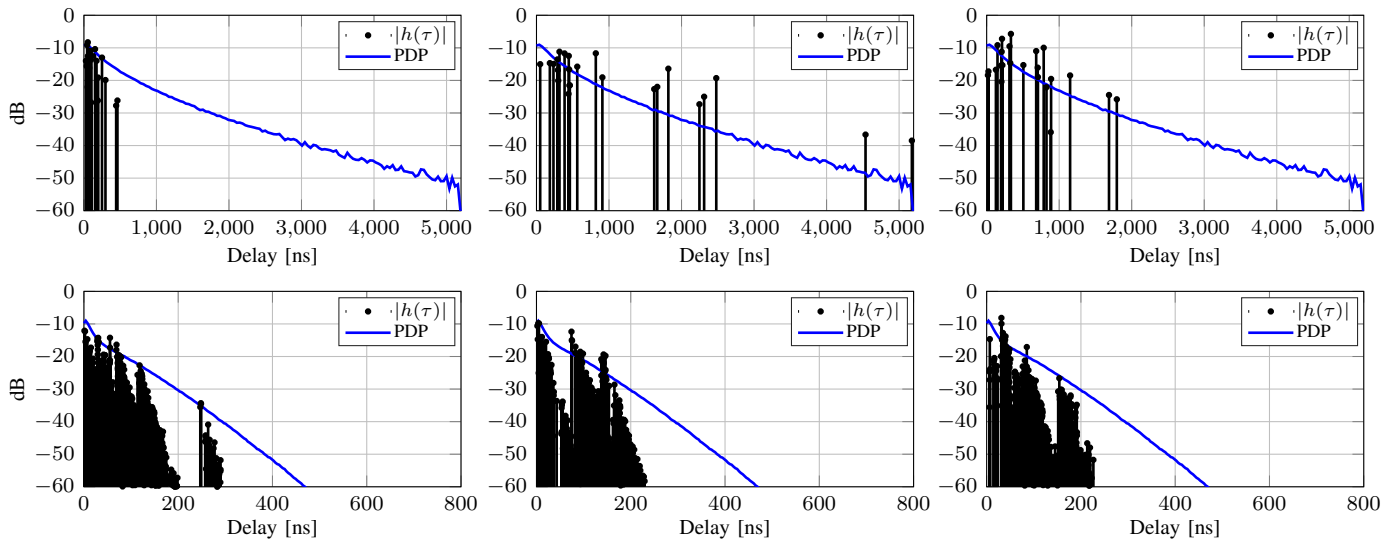

	\setlength\figureheight{26mm}
	\setlength\figurewidth{150pt}
	\pgfplotsset{local_axis_style/.style={
		legend style={at={(0.985,0.97)}, anchor=north east},
		y coord trafo/.code={\pgfmathparse{##1+100}},
    	y coord inv trafo/.code={\pgfmathparse{##1-100}}
	}} 
	\begin{minipage}[t]{0.35\linewidth}
		\centering
		\input{code/paper/tikz/chan_uma_nlos_1.tikz}
	\end{minipage}%
	\begin{minipage}[t]{0.33\linewidth}
		\centering
		\input{code/paper/tikz/chan_uma_nlos_2.tikz}
	\end{minipage}%
	\begin{minipage}[t]{0.32\linewidth}
		\centering
		\input{code/paper/tikz/chan_uma_nlos_3.tikz}
	\end{minipage} \\
	\begin{minipage}[t]{0.35\linewidth}
		\centering
		\input{code/paper/tikz/chan_ieee_1.tikz}
	\end{minipage}%
	\begin{minipage}[t]{0.33\linewidth}
		\centering
		\input{code/paper/tikz/chan_ieee_2.tikz}
	\end{minipage}%
	\begin{minipage}[t]{0.32\linewidth}
		\centering
		\input{code/paper/tikz/chan_ieee_3.tikz}
	\end{minipage}%
	\vspace{-2mm}%
	\caption{Three sample realizations of $h(\tau)$ for Scenario A (top) and
	Scenario B (bottom). An estimate of the PDP is also shown, which is
	obtained by averaging the magnitude-squared impulse responses of $50,000$
	channel realizations.}
	\label{fig:chan}
\end{figure*}

In our numerical evaluation we consider an OFDM system as described in Sec.
\ref{sec:model}. We use a random interleaver and a rate--1/2 non-systematic
convolutional channel code, decoded by the loopy BP implementation from the
Coded Modulation Library.%
\footnote{Available from http://iterativesolutions.com/Matlab.htm}
The pilot signals are chosen at random from a QPSK alphabet.
The first and last subcarriers are designated as pilots.
The other pilot
subcarriers are located equispaced%
\footnote{We have also conducted experiments with random pilot patterns (not
shown), but have seen no significant benefit in doing so for the setup
considered here.}
with spacing $\Delta_P$, i.e., the number of data subcarriers between two such
neighbour pilot subcarriers is $\Delta_P-1$.
The SNR is defined based on the realization of the CFR as
\begin{align}
	\mathrm{SNR} \triangleq \frac{\E{|x_i|^2} \lVert\bg\rVert^2}{N\beta},
\end{align}
where $\E{|x_i|^2}$ is calculated under the assumption that the symbols in the
respective alphabets $\bba_|D|$ and $\bba_|P|$ are equiprobable.

We asses how the receivers behave in two different scenarios.
The parameters considered in each scenario are listed in Table
\ref{tab:parameters}. Scenario A uses the channel model put forward by ITU
for the evaluation of IMT-Advanced radio interface technologies \cite{itu-m2135}.
Specifically we use the model with the parameter setting for urban macro (UMa)
environment with non line-of-sight (NLOS) conditions. The model generates
impulse responses $h(\tau)$ typical of macro-cellular communication in an urban
environment targeting continuous coverage for pedestrian up to fast vehicular
users \cite{itu-m2135}.  The channel model \cite{itu-m2135} is specified for
use with up to $100\;$MHz bandwidth, while the system we are simulating uses
$25.6\;$Mhz bandwidth. We are thus well within the specified bandwidth range.

Scenario B uses the standardized model proposed for the evaluation of IEEE 802.15.4a
UWB technologies \cite{molisch-comprehensive}. Specifically we use the model
with the setting proposed for outdoor environments with NLOS conditions. The
model generates impulse responses $h(\tau)$ typical of micro-cellular
communication in a suburban-like environment with a rather small range
\cite{molisch-comprehensive}. Note that this model is also used in
\cite{schniter-message}.

Since our signal model \eqref{obs} is not valid for CIRs longer than the cyclic
prefix duration $T_|CP|$, we drop realizations of the impulse response
$h(\tau)$ with component delays larger than $T_|CP|$. Fig. \ref{fig:chan} shows
3 impulse responses generated for each of scenarios A and B, along with an
estimate of the PDP.
An investigation of a few realizations has shown that in Scenario A
most (but not all) pairs of neighbouring multipath components adhere to a
separation by at least the reciprocal of the system bandwidth $1/(N\Delta_f)$. On the
other hand, in Scenario B, there are many pairs of neighbouring multipath
components that are not even separated by $10^{-2}/(N\Delta_f)$. In conclusion
the impulse responses in Scenario A generally show a specular behaviour, while in
Scenario B they show a dense behaviour.

We asses the performance of the considered receivers in terms of average coded
bit error rate (BER) and normalized mean squared error (MSE) of the CFR,
calculated as $\norm{\hat\bg-\bg}^2/\norm{\bg}^2$. These averages are obtained from
500 Monte Carlo trials ($\approx1.5\cdot10^6$ information bits) for SNR
$<20\db$, with one OFDM symbol transmitted in each trial. To get reliable BER
estimates we use $3,000$ trials ($\approx10^7$ information bits) for SNR
$=20\db$ and $15,000$ trials ($\approx4.5\cdot10^7$ information bits) for SNR
$>20\db$. The OFDM symbols and channel realizations are generated i.i.d.
according to the above.

\subsection{Evaluation and Comparison with Other Algorithms}
We evaluate our algorithm (Parametric BP-MF) and compare with the following
reference algorithms:

\textit{Turbo-GAMP \cite{schniter-message}}: The algorithm employs a
baud-spaced grid in the delay domain, i.e., the resolution of the grid is
$T_s=(N\Delta_f)\ii\approx39\,\text{ns}$ for Scenario A and
$T_s\approx3.9\,\text{ns}$ for Scenario B. For each channel tap a
large-tap and small-tap variance is provided along with tap-state transition
probabilities (see \cite{schniter-message} for more details). These are
estimated via the EM algorithm provided in \cite{schniter-message} from
$50,000$ channel realizations. Turbo-GAMP is provided with significant prior
information on the CIR via these statistical values. We also provide Turbo-GAMP
with the true noise variance, as \cite{schniter-message} does not give a way to
estimate this value.

\textit{LMMSE BP-MF \cite{badiu-message}}: The algorithm directly estimates the
CFR \bg via the BP-MF framework. It is an iterative receiver with LMMSE channel
estimation that requires prior knowledge of the noise variance and the
covariance matrix $\E{\bg\bg\h}$. We provide the true noise variance to the
receivers and show results using three different covariance matrices:
\begin{itemize}
	\item A receiver using the covariance matrix calculated from the robust PDP
		described in \cite{edfors-ofdm}, which assumes constant PDP within the
		interval $[0,T_|CP|)$. This is known to be an appropriate choice when
		no statistical information about the channel is available at the
		receiver \cite{edfors-ofdm}.
	\item A receiver using the true covariance matrix associated to the channel
		model. Due to the complex structure of the channel models, the true
		covariance matrix is not easy to obtain analytically. We therefore
		estimate it as the sample covariance matrix obtained from $50,000$
		channel realizations. We identify this estimate with its true
		counterpart. The use of the true covariance matrix corresponds to
		knowing the true PDP (of which an estimate is shown in Fig.
		\ref{fig:chan}). We refer to this receiver as LMMSE BP-MF with known
		PDP.
	\item An oracle receiver that calculates the channel covariance matrix
		conditioned on (i.e., knowing) the true delays and powers of the multipath
		components. This oracle receiver is thus provided with significant side
		information. We refer to it as LMMSE BP-MF with multipath oracle.
\end{itemize}
These three choices of the covariance matrix progressively assume stronger
prior knowledge or side information to be available at the receiver.

\textit{Perfect CSI}: This oracle receiver has perfect channel state
information (CSI), i.e., it knows the true CFR $\bg$, and thus provides a lower
bound on the achievable BER. The Perfect CSI trace is only shown in the BER
plots. To be specific, it is implemented by computing the messages
$n_{x_i\rightarrow f_{M_i}}(x_i)$ for all $i\in\cald$ (see
\eqref{message-x-fM}), followed by 5 iterations in the BP subgraph of Fig.
\ref{fig:fg}.


\setlength\figureheight{35mm}
\setlength\figurewidth{220pt}
\pgfplotsset{local_axis_style/.style={
	legend columns=2,
	legend style={at={(0.5,1.06)},anchor=south},
}} 
\begin{figure}[t!]
	\begin{minipage}[t]{\linewidth}
		\centering
%
\definecolor{mycolor1}{rgb}{1.00000,0.55000,0.00000}%
\definecolor{mycolor2}{rgb}{1.00000,0.00000,1.00000}%
\begin{tikzpicture}

\begin{axis}[%
width=0.951\figurewidth,
height=\figureheight,
at={(0\figurewidth,0\figureheight)},
scale only axis,
xmin=12,
xmax=22,
xlabel={SNR [dB]},
xmajorgrids,
ymode=log,
ymin=1e-06,
ymax=1,
yminorticks=true,
ylabel={BER},
ymajorgrids,
yminorgrids,
axis background/.style={fill=white},
legend style={legend cell align=left,align=left,draw=white!15!black},
global_axis_style, local_axis_style,
xtick = {1.2000e+01,1.4000e+01,1.6000e+01,1.8000e+01,2.0000e+01,2.2000e+01},
ytick = {1.0e-06,1.0e-05,1.0e-04,1.0e-03,1.0e-02,1.0e-01,1.0e+00,1.0e+01,1.0e+02},
minor ytick = {1.0e-06,2.0e-06,3.0e-06,4.0e-06,5.0e-06,6.0e-06,7.0e-06,8.0e-06,9.0e-06,1.0e-05,2.0e-05,3.0e-05,4.0e-05,5.0e-05,6.0e-05,7.0e-05,8.0e-05,9.0e-05,1.0e-04,2.0e-04,3.0e-04,4.0e-04,5.0e-04,6.0e-04,7.0e-04,8.0e-04,9.0e-04,1.0e-03,2.0e-03,3.0e-03,4.0e-03,5.0e-03,6.0e-03,7.0e-03,8.0e-03,9.0e-03,1.0e-02,2.0e-02,3.0e-02,4.0e-02,5.0e-02,6.0e-02,7.0e-02,8.0e-02,9.0e-02,1.0e-01,2.0e-01,3.0e-01,4.0e-01,5.0e-01,6.0e-01,7.0e-01,8.0e-01,9.0e-01,1.0e+00,2.0e+00,3.0e+00,4.0e+00,5.0e+00,6.0e+00,7.0e+00,8.0e+00,9.0e+00,1.0e+01,2.0e+01,3.0e+01,4.0e+01,5.0e+01,6.0e+01,7.0e+01,8.0e+01,9.0e+01,1.0e+02,2.0e+02,3.0e+02,4.0e+02,5.0e+02,6.0e+02,7.0e+02,8.0e+02,9.0e+02}
]
\addplot [color=red,solid,line width=1.0pt,mark=*,mark options={solid}]
  table[row sep=crcr]{%
12	0.425590588235294\\
14	0.234589411764706\\
16	0.0210382352941176\\
18	0.00105235294117647\\
20	3.74509803921569e-05\\
22	1.3921568627451e-06\\
};
\addlegendentry{Parametric BP-MF};

\addplot [color=mycolor1,solid,line width=1.0pt,mark=triangle,mark options={solid,rotate=180}]
  table[row sep=crcr]{%
12	0.468488823529411\\
14	0.412762352941177\\
16	0.137968235294118\\
18	0.00302941176470588\\
20	0.00013656862745098\\
22	4.80392156862745e-06\\
};
\addlegendentry{LMMSE BP-MF with robust PDP};

\addplot [color=mycolor2,solid,line width=1.0pt,mark=o,mark options={solid}]
  table[row sep=crcr]{%
12	0.43113\\
14	0.304289411764706\\
16	0.0589164705882353\\
18	0.00496764705882353\\
20	0.00084205882352941\\
22	0.000372450980392157\\
};
\addlegendentry{Turbo-GAMP};

\addplot [color=blue,solid,line width=1.0pt,mark=triangle,mark options={solid}]
  table[row sep=crcr]{%
12	0.442041176470588\\
14	0.372365882352942\\
16	0.0874158823529412\\
18	0.00429352941176471\\
20	0.000174901960784314\\
22	6.80392156862745e-06\\
};
\addlegendentry{LMMSE BP-MF with known PDP};

\addplot [color=black,dashed,line width=1.0pt]
  table[row sep=crcr]{%
12	0.393608235294117\\
14	0.186144705882353\\
16	0.0172229411764706\\
18	0.00084235294117647\\
20	3.24509803921569e-05\\
22	1e-06\\
};
\addlegendentry{Perfect CSI};

\addplot [color=green,solid,line width=1.0pt,mark=x,mark options={solid}]
  table[row sep=crcr]{%
12	0.405818235294118\\
14	0.208692352941176\\
16	0.0192552941176471\\
18	0.00107058823529412\\
20	3.14705882352941e-05\\
22	1.07843137254902e-06\\
};
\addlegendentry{LMMSE BP-MF with multipath oracle};

\end{axis}
\end{tikzpicture}%
%
\definecolor{mycolor1}{rgb}{1.00000,0.55000,0.00000}%
\definecolor{mycolor2}{rgb}{1.00000,0.00000,1.00000}%
\begin{tikzpicture}

\begin{axis}[%
width=0.951\figurewidth,
height=\figureheight,
at={(0\figurewidth,0\figureheight)},
scale only axis,
xmin=12,
xmax=22,
xlabel={SNR [dB]},
xmajorgrids,
ymode=log,
ymin=0.0001,
ymax=0.1,
yminorticks=true,
ylabel={MSE},
ymajorgrids,
yminorgrids,
axis background/.style={fill=white},
global_axis_style, local_axis_style,
xtick = {1.2000e+01,1.4000e+01,1.6000e+01,1.8000e+01,2.0000e+01,2.2000e+01},
ytick = {1.0e-06,1.0e-05,1.0e-04,1.0e-03,1.0e-02,1.0e-01,1.0e+00,1.0e+01,1.0e+02},
minor ytick = {1.0e-06,2.0e-06,3.0e-06,4.0e-06,5.0e-06,6.0e-06,7.0e-06,8.0e-06,9.0e-06,1.0e-05,2.0e-05,3.0e-05,4.0e-05,5.0e-05,6.0e-05,7.0e-05,8.0e-05,9.0e-05,1.0e-04,2.0e-04,3.0e-04,4.0e-04,5.0e-04,6.0e-04,7.0e-04,8.0e-04,9.0e-04,1.0e-03,2.0e-03,3.0e-03,4.0e-03,5.0e-03,6.0e-03,7.0e-03,8.0e-03,9.0e-03,1.0e-02,2.0e-02,3.0e-02,4.0e-02,5.0e-02,6.0e-02,7.0e-02,8.0e-02,9.0e-02,1.0e-01,2.0e-01,3.0e-01,4.0e-01,5.0e-01,6.0e-01,7.0e-01,8.0e-01,9.0e-01,1.0e+00,2.0e+00,3.0e+00,4.0e+00,5.0e+00,6.0e+00,7.0e+00,8.0e+00,9.0e+00,1.0e+01,2.0e+01,3.0e+01,4.0e+01,5.0e+01,6.0e+01,7.0e+01,8.0e+01,9.0e+01,1.0e+02,2.0e+02,3.0e+02,4.0e+02,5.0e+02,6.0e+02,7.0e+02,8.0e+02,9.0e+02}
]
\addplot [color=red,solid,line width=1.0pt,mark=*,mark options={solid},forget plot]
  table[row sep=crcr]{%
12	0.00881184934635146\\
14	0.00302620647109423\\
16	0.00093865672332665\\
18	0.000581677458560394\\
20	0.00039342945842756\\
22	0.000263020301697023\\
};
\addplot [color=mycolor1,solid,line width=1.0pt,mark=triangle,mark options={solid,rotate=180},forget plot]
  table[row sep=crcr]{%
12	0.0347722539683869\\
14	0.0251301479404021\\
16	0.00764149860639865\\
18	0.00232749087615526\\
20	0.00144412651913961\\
22	0.00090925939092951\\
};
\addplot [color=mycolor2,solid,line width=1.0pt,mark=o,mark options={solid},forget plot]
  table[row sep=crcr]{%
12	0.0146541082227458\\
14	0.0109000948402417\\
16	0.00528261714431797\\
18	0.00341051735600158\\
20	0.00247815760739706\\
22	0.00190463605457891\\
};
\addplot [color=blue,solid,line width=1.0pt,mark=triangle,mark options={solid},forget plot]
  table[row sep=crcr]{%
12	0.0173683258786742\\
14	0.0142953874626948\\
16	0.00492922267211933\\
18	0.00221130845353015\\
20	0.00122445555122432\\
22	0.000801292691728322\\
};
\addplot [color=green,solid,line width=1.0pt,mark=x,mark options={solid},forget plot]
  table[row sep=crcr]{%
12	0.00308344682311454\\
14	0.00136269341992626\\
16	0.000492561970542861\\
18	0.000293460404445723\\
20	0.000186320031947287\\
22	0.000118815213943331\\
};
\end{axis}
\end{tikzpicture}%
	\end{minipage}
	\caption{BER (top) and MSE of CFR (bottom) vs. SNR in Scenario A.}
	\label{fig:snr_uma_nlos}
\end{figure}

\subsection{Varying the Signal-to-Noise Ratio}
Fig. \ref{fig:snr_uma_nlos} shows performance results for varying SNR in
Scenario A. We first note that Parametric BP-MF performs very well in both BER
and MSE. Its BER is remarkably close to that of the two oracle estimators
(LMMSE BP-MF with multipath oracle and Perfect CSI), indicating that there is
very little margin for improvement of the algorithm in this scenario.
The robust and known PDP versions of LMMSE BP-MF show higher BER than
Parametric BP-MF, corresponding to a decrease in SNR of about $1$\db. They show
almost the same BER performance because the delay spread in Scenario A is
relatively large (cf. Fig.~\ref{fig:chan}) and the robust PDP
assumption is therefore realistic. Turbo-GAMP does not perform
well and shows a BER floor at high SNR. The reason is discussed below.

Fig. \ref{fig:snr_ieee} shows the corresponding results for Scenario B. We here
observe that Parametric BP-MF has a BER loss compared to the Perfect CSI trace
corresponding to about $0.5\db$ SNR difference. Parametric BP-MF is among the
best performing algorithms, even though the impulse responses generated in Scenario B are
dense and thus composed of a very large number of multipath components that the
algorithm cannot resolve individually (cf. Fig.~\ref{fig:chan}).
Instead, the algorithm estimates a virtual CIR with significantly fewer
components that approximates the true CIR within the system
bandwidth. We have observed that the estimated virtual CIR approximately
``recovers the support'' of the true CIR, in the sense that an estimated
multipath component is located wherever the CIR contains significant power.
Parametric BP-MF has a BER and MSE performance equivalent to that of LMMSE
BP-MF with both known PDP and multipath oracle. We stress that Parametric BP-MF
achieves this performance without using prior knowledge of the channel.

\begin{figure}[t!]
	\begin{minipage}[t]{\linewidth}
		\centering
%
\definecolor{mycolor1}{rgb}{1.00000,0.55000,0.00000}%
\definecolor{mycolor2}{rgb}{1.00000,0.00000,1.00000}%
\begin{tikzpicture}

\begin{axis}[%
width=0.951\figurewidth,
height=\figureheight,
at={(0\figurewidth,0\figureheight)},
scale only axis,
xmin=12,
xmax=22,
xlabel={SNR [dB]},
xmajorgrids,
ymode=log,
ymin=1e-06,
ymax=1,
yminorticks=true,
ylabel={BER},
ymajorgrids,
yminorgrids,
axis background/.style={fill=white},
legend style={legend cell align=left,align=left,draw=white!15!black},
global_axis_style, local_axis_style,
xtick = {1.2000e+01,1.4000e+01,1.6000e+01,1.8000e+01,2.0000e+01,2.2000e+01},
ytick = {1.0e-06,1.0e-05,1.0e-04,1.0e-03,1.0e-02,1.0e-01,1.0e+00,1.0e+01,1.0e+02},
minor ytick = {1.0e-06,2.0e-06,3.0e-06,4.0e-06,5.0e-06,6.0e-06,7.0e-06,8.0e-06,9.0e-06,1.0e-05,2.0e-05,3.0e-05,4.0e-05,5.0e-05,6.0e-05,7.0e-05,8.0e-05,9.0e-05,1.0e-04,2.0e-04,3.0e-04,4.0e-04,5.0e-04,6.0e-04,7.0e-04,8.0e-04,9.0e-04,1.0e-03,2.0e-03,3.0e-03,4.0e-03,5.0e-03,6.0e-03,7.0e-03,8.0e-03,9.0e-03,1.0e-02,2.0e-02,3.0e-02,4.0e-02,5.0e-02,6.0e-02,7.0e-02,8.0e-02,9.0e-02,1.0e-01,2.0e-01,3.0e-01,4.0e-01,5.0e-01,6.0e-01,7.0e-01,8.0e-01,9.0e-01,1.0e+00,2.0e+00,3.0e+00,4.0e+00,5.0e+00,6.0e+00,7.0e+00,8.0e+00,9.0e+00,1.0e+01,2.0e+01,3.0e+01,4.0e+01,5.0e+01,6.0e+01,7.0e+01,8.0e+01,9.0e+01,1.0e+02,2.0e+02,3.0e+02,4.0e+02,5.0e+02,6.0e+02,7.0e+02,8.0e+02,9.0e+02}
]
\addplot [color=red,solid,line width=1.0pt,mark=*,mark options={solid}]
  table[row sep=crcr]{%
12	0.45292091503268\\
14	0.331612418300654\\
16	0.0552300653594773\\
18	0.00242679738562091\\
20	0.000110348583877995\\
22	3.89978213507625e-06\\
};
\addlegendentry{Parametric BP-MF};

\addplot [color=mycolor1,solid,line width=1.0pt,mark=triangle,mark options={solid,rotate=180}]
  table[row sep=crcr]{%
12	0.474619607843137\\
14	0.431030718954248\\
16	0.243421568627451\\
18	0.0158503267973856\\
20	0.000467864923747275\\
22	2.11111111111111e-05\\
};
\addlegendentry{LMMSE BP-MF with robust PDP};

\addplot [color=mycolor2,solid,line width=1.0pt,mark=o,mark options={solid}]
  table[row sep=crcr]{%
12	0.436311764705882\\
14	0.325567973856209\\
16	0.0675222222222224\\
18	0.00572418300653594\\
20	0.000605228758169933\\
22	0.000233616557734204\\
};
\addlegendentry{Turbo-GAMP};

\addplot [color=blue,solid,line width=1.0pt,mark=triangle,mark options={solid}]
  table[row sep=crcr]{%
12	0.437436601307189\\
14	0.340967320261438\\
16	0.0542483660130721\\
18	0.0023235294117647\\
20	8.89978213507624e-05\\
22	3.79084967320261e-06\\
};
\addlegendentry{LMMSE BP-MF with known PDP};

\addplot [color=black,dashed,line width=1.0pt]
  table[row sep=crcr]{%
12	0.399599346405229\\
14	0.201505882352941\\
16	0.0201895424836602\\
18	0.0010640522875817\\
20	4.18300653594771e-05\\
22	1.74291938997821e-06\\
};
\addlegendentry{Perfect CSI};

\addplot [color=green,solid,line width=1.0pt,mark=x,mark options={solid}]
  table[row sep=crcr]{%
12	0.425417647058824\\
14	0.294364705882353\\
16	0.0353666666666668\\
18	0.00175359477124183\\
20	7.18954248366012e-05\\
22	2.2875816993464e-06\\
};
\addlegendentry{LMMSE BP-MF with multipath oracle};

\end{axis}
\end{tikzpicture}%
%
\definecolor{mycolor1}{rgb}{1.00000,0.55000,0.00000}%
\definecolor{mycolor2}{rgb}{1.00000,0.00000,1.00000}%
\begin{tikzpicture}

\begin{axis}[%
width=0.951\figurewidth,
height=\figureheight,
at={(0\figurewidth,0\figureheight)},
scale only axis,
xmin=12,
xmax=22,
xlabel={SNR [dB]},
xmajorgrids,
ymode=log,
ymin=0.0001,
ymax=0.1,
yminorticks=true,
ylabel={MSE},
ymajorgrids,
yminorgrids,
axis background/.style={fill=white},
global_axis_style, local_axis_style,
xtick = {1.2000e+01,1.4000e+01,1.6000e+01,1.8000e+01,2.0000e+01,2.2000e+01},
ytick = {1.0e-06,1.0e-05,1.0e-04,1.0e-03,1.0e-02,1.0e-01,1.0e+00,1.0e+01,1.0e+02},
minor ytick = {1.0e-06,2.0e-06,3.0e-06,4.0e-06,5.0e-06,6.0e-06,7.0e-06,8.0e-06,9.0e-06,1.0e-05,2.0e-05,3.0e-05,4.0e-05,5.0e-05,6.0e-05,7.0e-05,8.0e-05,9.0e-05,1.0e-04,2.0e-04,3.0e-04,4.0e-04,5.0e-04,6.0e-04,7.0e-04,8.0e-04,9.0e-04,1.0e-03,2.0e-03,3.0e-03,4.0e-03,5.0e-03,6.0e-03,7.0e-03,8.0e-03,9.0e-03,1.0e-02,2.0e-02,3.0e-02,4.0e-02,5.0e-02,6.0e-02,7.0e-02,8.0e-02,9.0e-02,1.0e-01,2.0e-01,3.0e-01,4.0e-01,5.0e-01,6.0e-01,7.0e-01,8.0e-01,9.0e-01,1.0e+00,2.0e+00,3.0e+00,4.0e+00,5.0e+00,6.0e+00,7.0e+00,8.0e+00,9.0e+00,1.0e+01,2.0e+01,3.0e+01,4.0e+01,5.0e+01,6.0e+01,7.0e+01,8.0e+01,9.0e+01,1.0e+02,2.0e+02,3.0e+02,4.0e+02,5.0e+02,6.0e+02,7.0e+02,8.0e+02,9.0e+02}
]
\addplot [color=red,solid,line width=1.0pt,mark=*,mark options={solid},forget plot]
  table[row sep=crcr]{%
12	0.0185282160728269\\
14	0.00979765977172056\\
16	0.00373527730710807\\
18	0.00205250832496596\\
20	0.00135272811614197\\
22	0.000900235954085472\\
};
\addplot [color=mycolor1,solid,line width=1.0pt,mark=triangle,mark options={solid,rotate=180},forget plot]
  table[row sep=crcr]{%
12	0.0421036997567546\\
14	0.0306411466254317\\
16	0.0144192251555703\\
18	0.00392151827578036\\
20	0.00222879075878104\\
22	0.00140223309715671\\
};
\addplot [color=mycolor2,solid,line width=1.0pt,mark=o,mark options={solid},forget plot]
  table[row sep=crcr]{%
12	0.0144649071648021\\
14	0.0108889141745827\\
16	0.00538288557530982\\
18	0.00338814432675101\\
20	0.00248981224258746\\
22	0.00188184210703912\\
};
\addplot [color=blue,solid,line width=1.0pt,mark=triangle,mark options={solid},forget plot]
  table[row sep=crcr]{%
12	0.0121076137499132\\
14	0.00894110393933572\\
16	0.00293365522238593\\
18	0.00158070866713449\\
20	0.00102648561368367\\
22	0.000678486061723424\\
};
\addplot [color=green,solid,line width=1.0pt,mark=x,mark options={solid},forget plot]
  table[row sep=crcr]{%
12	0.00779244001470748\\
14	0.00526192228885093\\
16	0.00185096210462356\\
18	0.0010856205053843\\
20	0.000710691743751657\\
22	0.000469579853533443\\
};
\end{axis}
\end{tikzpicture}%
	\end{minipage}
	\caption{BER (top) and MSE of CFR (bottom) vs. SNR in Scenario B.}
	\label{fig:snr_ieee}
\end{figure}

In Scenario B we observe a significant difference between the LMMSE BP-MF
algorithms with known and robust PDP. To explain this difference observe in
Fig. \ref{fig:chan} that most of the mass of the PDP is located at small
delays. This significantly deviates from the evenly distributed mass on $[0,
T_|CP|)$ that underlies the robust PDP assumption.

In both Fig.~\ref{fig:snr_uma_nlos} and Fig.~\ref{fig:snr_ieee} an error floor
is observed for Turbo-GAMP at high SNR.%
\footnote{In \cite{schniter-message}, where Turbo-GAMP is introduced,
such an error floor is not observed even though the setup in the numerical
investigation is almost identical to that in Scenario B.
The reason is an error in the signal model in \cite{schniter-message} that
invalidates the numerical results obtained in that paper. Specifically  the
error occurs when the ``uniformly sampled channel taps'' are defined as rate
$1/T$ samples of the compound CIR $x(\tau)\triangleq (g_r\ast h\ast g_t)(\tau)$
(notation as in \cite{schniter-message}).  However, since $(g_r\ast g_t)(\tau)$
is a raised-cosine filter with design parameter $0.5$, $x(\tau)$ has bandwidth
$1.5/T$, leading to aliasing in the sampling operation.}
We conjecture that this error floor is caused by the restriction of the delays
to the baud-spaced grid. If the delays are generated to be located on that
grid, the performance of Turbo-GAMP is very close to that of the Perfect CSI
trace (not shown here).
We have also conducted experiments with random pilot patterns (not shown) as
used in \cite{schniter-message} (where Turbo-GAMP is introduced) but did
not see an improvement of Turbo-GAMP in that case. We note that such error
floors in BER and MSE have previously been observed for other grid-based sparse
channel estimation algorithms, see for example \cite{taubock-sparsechannel,
barbu-sparse}. In conclusion, the grid-based approximation is of insufficient
accuracy for communication with large modulation order in the high-SNR regime.

\begin{figure}[t!]
	\begin{minipage}[t]{\linewidth}
		\centering
%
\definecolor{mycolor1}{rgb}{1.00000,0.55000,0.00000}%
\definecolor{mycolor2}{rgb}{1.00000,0.00000,1.00000}%
\begin{tikzpicture}

\begin{axis}[%
width=0.951\figurewidth,
height=\figureheight,
at={(0\figurewidth,0\figureheight)},
scale only axis,
xmin=38,
xmax=342,
xlabel={Number of pilots},
xmajorgrids,
ymode=log,
ymin=1e-05,
ymax=1,
yminorticks=true,
ylabel={BER},
ymajorgrids,
yminorgrids,
axis background/.style={fill=white},
legend style={legend cell align=left,align=left,draw=white!15!black},
global_axis_style, local_axis_style,
xtick = {3.8000e+01,7.0000e+01,1.0400e+02,1.4800e+02,2.0600e+02,2.5700e+02,3.4200e+02},
ytick = {1.0e-06,1.0e-05,1.0e-04,1.0e-03,1.0e-02,1.0e-01,1.0e+00,1.0e+01,1.0e+02},
minor ytick = {1.0e-06,2.0e-06,3.0e-06,4.0e-06,5.0e-06,6.0e-06,7.0e-06,8.0e-06,9.0e-06,1.0e-05,2.0e-05,3.0e-05,4.0e-05,5.0e-05,6.0e-05,7.0e-05,8.0e-05,9.0e-05,1.0e-04,2.0e-04,3.0e-04,4.0e-04,5.0e-04,6.0e-04,7.0e-04,8.0e-04,9.0e-04,1.0e-03,2.0e-03,3.0e-03,4.0e-03,5.0e-03,6.0e-03,7.0e-03,8.0e-03,9.0e-03,1.0e-02,2.0e-02,3.0e-02,4.0e-02,5.0e-02,6.0e-02,7.0e-02,8.0e-02,9.0e-02,1.0e-01,2.0e-01,3.0e-01,4.0e-01,5.0e-01,6.0e-01,7.0e-01,8.0e-01,9.0e-01,1.0e+00,2.0e+00,3.0e+00,4.0e+00,5.0e+00,6.0e+00,7.0e+00,8.0e+00,9.0e+00,1.0e+01,2.0e+01,3.0e+01,4.0e+01,5.0e+01,6.0e+01,7.0e+01,8.0e+01,9.0e+01,1.0e+02,2.0e+02,3.0e+02,4.0e+02,5.0e+02,6.0e+02,7.0e+02,8.0e+02,9.0e+02}
]
\addplot [color=red,solid,line width=1.0pt,mark=*,mark options={solid}]
  table[row sep=crcr]{%
38	0.0280995934959349\\
70	0.000825630252100841\\
104	0.000218500363108206\\
148	3.23226544622426e-05\\
206	2.96160130718954e-05\\
257	4.56427015250545e-05\\
342	4.66911764705882e-05\\
};
\addlegendentry{Parametric BP-MF};

\addplot [color=mycolor1,solid,line width=1.0pt,mark=triangle,mark options={solid,rotate=180}]
  table[row sep=crcr]{%
38	0.500162178184282\\
70	0.499772408963585\\
104	0.499400417574437\\
148	0.000105644546147979\\
206	0.000133476307189542\\
257	0.000116666666666666\\
342	0.000127083333333333\\
};
\addlegendentry{LMMSE BP-MF with robust PDP};

\addplot [color=mycolor2,solid,line width=1.0pt,mark=o,mark options={solid}]
  table[row sep=crcr]{%
38	0.0606689532520323\\
70	0.00554858193277312\\
104	0.00155492011619463\\
148	0.000760106788710908\\
206	0.000729370915032681\\
257	0.000769063180827885\\
342	0.000690318627450983\\
};
\addlegendentry{Turbo-GAMP};

\addplot [color=blue,solid,line width=1.0pt,mark=triangle,mark options={solid}]
  table[row sep=crcr]{%
38	0.453708587398374\\
70	0.0661092436974789\\
104	0.0017695170660857\\
148	9.82074752097635e-05\\
206	0.000119791666666666\\
257	9.99999999999998e-05\\
342	0.000107598039215686\\
};
\addlegendentry{LMMSE BP-MF with known PDP};

\addplot [color=black,dashed,line width=1.0pt]
  table[row sep=crcr]{%
38	3.92953929539296e-05\\
70	3.35259103641456e-05\\
104	3.73093681917211e-05\\
148	2.63157894736842e-05\\
206	2.7062908496732e-05\\
257	3.18082788671024e-05\\
342	3.30882352941176e-05\\
};
\addlegendentry{Perfect CSI};

\addplot [color=green,solid,line width=1.0pt,mark=x,mark options={solid}]
  table[row sep=crcr]{%
38	0.000348492547425474\\
70	3.1687675070028e-05\\
104	4.42992011619462e-05\\
148	3.09877955758963e-05\\
206	2.73692810457516e-05\\
257	3.89978213507625e-05\\
342	3.8235294117647e-05\\
};
\addlegendentry{LMMSE BP-MF with multipath oracle};

\end{axis}
\end{tikzpicture}%
		\pgfplotsset{local_axis_style/.style={
			every axis legend/.code={\let\addlegendentry\relax}
		}}
%
\definecolor{mycolor1}{rgb}{1.00000,0.55000,0.00000}%
\definecolor{mycolor2}{rgb}{1.00000,0.00000,1.00000}%
\begin{tikzpicture}

\begin{axis}[%
width=0.951\figurewidth,
height=\figureheight,
at={(0\figurewidth,0\figureheight)},
scale only axis,
xmin=38,
xmax=342,
xlabel={Number of pilots},
xmajorgrids,
ymode=log,
ymin=1e-05,
ymax=1,
yminorticks=true,
ylabel={BER},
ymajorgrids,
yminorgrids,
axis background/.style={fill=white},
legend style={legend cell align=left,align=left,draw=white!15!black},
global_axis_style, local_axis_style,
xtick = {3.8000e+01,7.0000e+01,1.0400e+02,1.4800e+02,2.0600e+02,2.5700e+02,3.4200e+02},
ytick = {1.0e-06,1.0e-05,1.0e-04,1.0e-03,1.0e-02,1.0e-01,1.0e+00,1.0e+01,1.0e+02},
minor ytick = {1.0e-06,2.0e-06,3.0e-06,4.0e-06,5.0e-06,6.0e-06,7.0e-06,8.0e-06,9.0e-06,1.0e-05,2.0e-05,3.0e-05,4.0e-05,5.0e-05,6.0e-05,7.0e-05,8.0e-05,9.0e-05,1.0e-04,2.0e-04,3.0e-04,4.0e-04,5.0e-04,6.0e-04,7.0e-04,8.0e-04,9.0e-04,1.0e-03,2.0e-03,3.0e-03,4.0e-03,5.0e-03,6.0e-03,7.0e-03,8.0e-03,9.0e-03,1.0e-02,2.0e-02,3.0e-02,4.0e-02,5.0e-02,6.0e-02,7.0e-02,8.0e-02,9.0e-02,1.0e-01,2.0e-01,3.0e-01,4.0e-01,5.0e-01,6.0e-01,7.0e-01,8.0e-01,9.0e-01,1.0e+00,2.0e+00,3.0e+00,4.0e+00,5.0e+00,6.0e+00,7.0e+00,8.0e+00,9.0e+00,1.0e+01,2.0e+01,3.0e+01,4.0e+01,5.0e+01,6.0e+01,7.0e+01,8.0e+01,9.0e+01,1.0e+02,2.0e+02,3.0e+02,4.0e+02,5.0e+02,6.0e+02,7.0e+02,8.0e+02,9.0e+02}
]
\addplot [color=red,solid,line width=1.0pt,mark=*,mark options={solid}]
  table[row sep=crcr]{%
38	0.0913270663956641\\
70	0.000850402661064427\\
104	9.36819172113289e-05\\
148	9.00076277650649e-05\\
206	0.000119893790849673\\
257	9.61873638344226e-05\\
342	0.00010906862745098\\
};
\addlegendentry{Parametric BP-MF};

\addplot [color=mycolor1,solid,line width=1.0pt,mark=triangle,mark options={solid,rotate=180}]
  table[row sep=crcr]{%
38	0.500023035230352\\
70	0.499916491596638\\
104	0.499960148874366\\
148	0.499783466819223\\
206	0.0693210784313727\\
257	0.000480610021786491\\
342	0.000446936274509805\\
};
\addlegendentry{LMMSE BP-MF with robust PDP};

\addplot [color=mycolor2,solid,line width=1.0pt,mark=o,mark options={solid}]
  table[row sep=crcr]{%
38	0.0802613482384821\\
70	0.00261545868347339\\
104	0.00106290849673203\\
148	0.000629099923722349\\
206	0.000670547385620917\\
257	0.00067342047930283\\
342	0.00067230392156863\\
};
\addlegendentry{Turbo-GAMP};

\addplot [color=blue,solid,line width=1.0pt,mark=triangle,mark options={solid}]
  table[row sep=crcr]{%
38	0.466733485772357\\
70	0.00467191876750701\\
104	7.54357298474945e-05\\
148	8.6003051106026e-05\\
206	9.50776143790849e-05\\
257	8.87799564270152e-05\\
342	9.80392156862744e-05\\
};
\addlegendentry{LMMSE BP-MF with known PDP};

\addplot [color=black,dashed,line width=1.0pt]
  table[row sep=crcr]{%
38	3.92953929539296e-05\\
70	3.88655462184874e-05\\
104	3.11365286855483e-05\\
148	3.47063310450038e-05\\
206	4.17687908496732e-05\\
257	3.81263616557734e-05\\
342	4.01960784313725e-05\\
};
\addlegendentry{Perfect CSI};

\addplot [color=green,solid,line width=1.0pt,mark=x,mark options={solid}]
  table[row sep=crcr]{%
38	0.109334857723577\\
70	0.000878764005602242\\
104	5.82788671023965e-05\\
148	7.06521739130435e-05\\
206	7.18954248366013e-05\\
257	7.17864923747276e-05\\
342	8.28431372549019e-05\\
};
\addlegendentry{LMMSE BP-MF with multipath oracle};

\end{axis}
\end{tikzpicture}%
	\end{minipage}
	\caption{BER vs. number of pilot subcarriers in Scenario A (top) and
	Scenario B (bottom) at 20\db SNR.}
	\label{fig:pilotspacing_uma_nlos}
\end{figure}
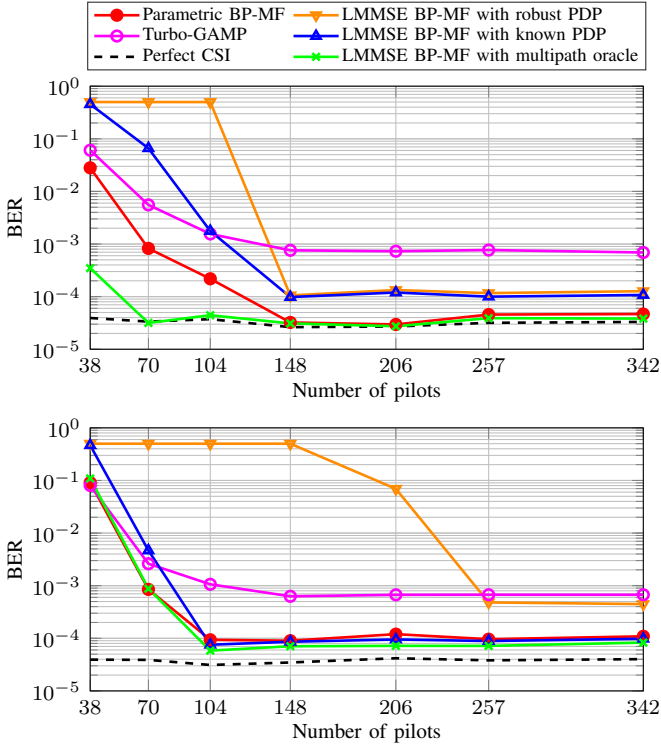

\subsection{Varying the Number of Pilots}
\label{sec:pilotspacing}
We now investigate if our receiver design improves the trade-off between the
number of pilots and estimator performance. Fig.
\ref{fig:pilotspacing_uma_nlos} shows the BER performance for varying number of
pilot subcarriers.

The first observation is that LMMSE BP-MF with robust PDP
shows a point at which the BER performance quickly transitions between high
($50\,\%$) BER and low ($<10^{-3}$) BER. Under the robust PDP
assumption, the channel coherence bandwidth is approximately $1/T_|CP|$. As
a rule of thumb there should be at least one pilot subcarrier per coherence
interval, which gives
the criterion $P>N\Delta_fT_|CP|$, where $P$ is the needed number of pilot
subcarriers. For Scenario A we have $N\Delta_fT_|CP|\approx133$ and for
Scenario B we have $N\Delta_fT_|CP|\approx205$, which exactly are the
respective numbers of pilots at which LMMSE BP-MF with robust PDP transitions
between low and high BER.

All algorithms except LMMSE BP-MF with robust PDP can operate significantly
below the above-mentioned limit. Due to the iterative processing, the
number of pilots can be decreased significantly without incurring an increase
in BER.


\section{Conclusions} \label{sec:conclusions}
In this paper we proposed an iterative OFDM receiver that employs sparsity-based
parametric channel estimation. The iterative receiver is derived using the BP-MF
framework for approximate Bayesian inference. Unlike state-of-the-art sparse channel
estimators, our scheme does not restrict multipath delays of the estimated
channel impulse response to a grid. As a result it can truly exploit parsimony of the
channel impulse response, without resorting to approximate sparsity (as in
\cite{schniter-message,prasad-joint,prasad-joint-mimo}).

We have presented a numerical evaluation that compares our algorithm with
state-of-the-art methods, i.e., Turbo-GAMP \cite{schniter-message} and LMMSE
BP-MF \cite{badiu-message}. This study demonstrated that restricting the
multipath delays to a baud-spaced grid (e.g., as in Turbo-GAMP) is not a viable
approach because the resulting equivalent vector of channel taps is only
approximately sparse.

The numerical evaluation also shows that our proposed scheme can effectively
exploit the structure of wireless channel impulse responses. We have showed
numerically that parametric channel estimation works well with both specular
and dense channels.
Our analysis of the channel covariance matrix in Sec.~\ref{sec:chan} shows that
for dense channels a virtual channel impulse response can be estimated, with a
number of virtual components given by the (effective) rank of the channel
covariance matrix. The corresponding virtual frequency response approximates
the actual channel frequency response well within the system bandwidth.

\appendices

\section{The Region-Based Free Energy Approximation}
\label{app:rbfe}
At the heart of the derivation of our algorithm lies the RBFE as defined by
\cite[Eq. (17)]{riegler-merging}, \cite{yedidia-constructing}. In this paper we
use the RBFE of the probability distribution corresponding to the factor graph
depicted in Fig. \ref{fig:fg}. For convenience, we give here the complete
expression of the RBFE:
\begin{align}
	F_|BP-MF| = F_|BP| + F_|MF|
	\label{rbfe}
\end{align}
with
\vspace{-1em}
\begin{multline*}
	F_|BP| =
		\sum_{k\in\calk} \sum_{u_k\in\{0,1\}} b_{u_k}(u_k)
			\ln\frac{b_{u_k}(u_k)}{p(u_k)}
		\\
		+ \sum_{i\in\cald}\;\;\;
			\sum_{\mathclap{\substack{x_i\in\bba_|D|\\\bc^{(i)}\in\{0,1\}^Q}}}
			b_{M_i}(x_i,\bc^{(i)})
			\ln\frac{b_{M_i}(x_i,\bc^{(i)})}{p(x_i|\bc^{(i)})}
		\\
		+ \sum_{\mathclap{\substack{\bc\in\{0,1\}^{K/R}\\\bu\in\{0,1\}^K}}} b_C(\bc,\bu)
			\ln\frac{b_C(\bc,\bu)}{p(\bc|\bu)}
		- \sum_{k\in\calk} \sum_{u_k\in\{0,1\}} q(u_k) \ln q(u_k)
		\\
		- \sum_{i\in\cald}\sum_{m\in\{1,\ldots,Q\}} \sum_{[\bc^{(i)}]_m\in\{0,1\}} 
			q([\bc^{(i)}]_m) \ln q([\bc^{(i)}]_m),
\end{multline*}
\vspace{-1.3em}
\begin{multline*}
	F_|MF| = \sum_{l\in\call}
			\int q(\alpha_l)\ln q(\alpha_l) \,\dd\alpha_l
		\\
		- \left<
			\ln p(\by|\bx_\cald,\balpha,\hat\btau;\hat\beta) \right>_{\bx_\cald,\balpha}
		\\
		- \sum_{l\in\call}
			\left< \ln p(\alpha_l|\hat z_l;\hat\eta)
			p(\hat z_l;\hat\rho)p(\hat\tau_l) \right>_{\alpha_l},
\end{multline*}
where $b_C(\bc,\bu)$, $b_{M_i}(x_i,\bc^{(i)})$ for $i\in\cald$ and
$b_{u_k}(u_k)$ for $k\in\calk$ are factor beliefs. With abuse of notation we
let $q(\cdot)$ denote variable beliefs and $\left<\cdot\right>_{a}$ denote
expectation with respect to the belief density $q(a)$.

\section{Efficient Calculation of
\texorpdfstring{$\hat\bmu_{\hat\cala}$ When $\hat L$}{u When L} is Large}
\label{app:bmu}

In this appendix we present a computationally efficient method for evaluating
$\hat\bmu_{\hat\cala}$ as defined by \eqref{naive-bmu-update}.  We here present
an iterative approach but note that an alternative (non-iterative) fast method
can most likely be obtained by extending the approach of
\cite{hansen-superfast}.

Direct evaluation and inversion of \bQ has time complexity $\calo(\hat L^2N)$,
where $\hat L\triangleq|\hat\cala|$. The iterative method presented here has
complexity $\calo(\hat LN\sqrt{N})$ provided Conjecture \ref{conj:eigvals}
(below) holds. It is thus beneficial to use it when $\hat L$ grows faster than
$\sqrt{N}$.

We first use the Woodbury matrix identity to write $\hat\bmu$ as
{\small
\begin{align*}
	\hat\bmu &= \hat\beta\ii\hat\eta \left(
	\bI - \hat\beta\ii\hat\eta \bPsi\h(\hat\btau_{\hat\cala})
			\bC\ii
			\bPsi(\hat\btau_{\hat\cala})
		\right)
		\bPsi\h(\hat\btau_{\hat\cala}) \left<\bX\right>_{\bx_{\cald}}\h \by,
\end{align*}
}
where
\begin{align*}
	\bC &= \left<\bX\h\bX\right>_{\bx_{\cald}}\ii
		+ \hat\beta\ii\hat\eta \bPsi(\hat\btau_{\hat\cala})
		\bPsi\h(\hat\btau_{\hat\cala}).
\end{align*}
We immediately recognize that the computationally dominating part is to solve a
system of $N$ linear equations of the form $\bC\bz=\ba$. Since \bC is Hermitian
and positive-definite, we can solve this system via the conjugate-gradient (CG)
method (Alg. 2.1 in \cite{ng-iterative}), which is an iterative method for
solving systems of linear equations. In the following we show that the number
of iterations of the CG method is $\calo(\sqrt{N})$.

\begin{figure}[t!]
	\setlength\figureheight{24mm}
	\setlength\figurewidth{105pt}
	\pgfplotsset{local_axis_style/.style={
		every x tick label/.append style={font=\footnotesize, rotate=35, yshift=2pt},
		every axis x label/.append style={yshift=3pt},
		scale ticks above exponent=1,
		scaled x ticks = false,
		scaled y ticks = true
	}} 
	\begin{minipage}[t]{0.5\linewidth}
		\vspace{0pt}
		\centering
%
\begin{tikzpicture}

\begin{axis}[%
width=0.951\figurewidth,
height=\figureheight,
at={(0\figurewidth,0\figureheight)},
scale only axis,
xmin=256,
xmax=2048,
xlabel={Number of subcarriers ($N$)},
xmajorgrids,
ymin=5000,
ymax=20000,
ymajorgrids,
axis background/.style={fill=white},
global_axis_style, local_axis_style,
xtick = {2.5600e+02,5.1200e+02,1.0240e+03,1.5360e+03,2.0480e+03}
]
\addplot [color=blue,line width=1.0pt,only marks,mark=x,mark options={solid},forget plot]
  table[row sep=crcr]{%
256	5794.12187544405\\
512	7989.69816826101\\
768	10230.0458536721\\
1024	11858.4721413252\\
1280	13919.3640134179\\
1536	16022.0527435668\\
1792	17877.05116011\\
2048	19483.8796733894\\
};
\addplot [color=black,dashed,line width=1.0pt,forget plot]
  table[row sep=crcr]{%
256	6034.25311670499\\
512	7994.99099868879\\
768	9955.72888067259\\
1024	11916.4667626564\\
1280	13877.2046446402\\
1536	15837.942526624\\
1792	17798.6804086078\\
2048	19759.4182905916\\
};
\end{axis}
\end{tikzpicture}%
	\end{minipage}%
	\begin{minipage}[t]{0.5\linewidth}
		\vspace{0pt}
		\centering
%
\begin{tikzpicture}

\begin{axis}[%
width=0.951\figurewidth,
height=\figureheight,
at={(0\figurewidth,0\figureheight)},
scale only axis,
xmin=256,
xmax=2048,
xlabel={Number of subcarriers ($N$)},
xmajorgrids,
ymin=3500,
ymax=7000,
ymajorgrids,
axis background/.style={fill=white},
global_axis_style, local_axis_style,
xtick = {2.5600e+02,5.1200e+02,1.0240e+03,1.5360e+03,2.0480e+03}
]
\addplot [color=blue,line width=1.0pt,only marks,mark=x,mark options={solid},forget plot]
  table[row sep=crcr]{%
256	3790.3809689475\\
512	4496.66084478752\\
768	4943.77355141123\\
1024	5352.90489857458\\
1280	5639.98623793054\\
1536	6066.93832769962\\
1792	6226.56197409627\\
2048	6509.52171322977\\
};
\addplot [color=black,dashed,line width=1.0pt,forget plot]
  table[row sep=crcr]{%
256	4072.5049593871\\
512	4445.60098944353\\
768	4818.69701949997\\
1024	5191.79304955641\\
1280	5564.88907961285\\
1536	5937.98510966929\\
1792	6311.08113972572\\
2048	6684.17716978216\\
};
\end{axis}
\end{tikzpicture}%
	\end{minipage}%
	\caption{Average of the largest eigenvalue of the matrix \bT encountered
	during one execution of Parametric BP-MF for Scenario A (left) and B
	(right). Average obtained from 1000 Monte Carlo trials. Both plots were
	generated at $20\db$ SNR. A dashed line depicts the least-squares linear fit.}
	\label{fig:saveeig}
\end{figure}
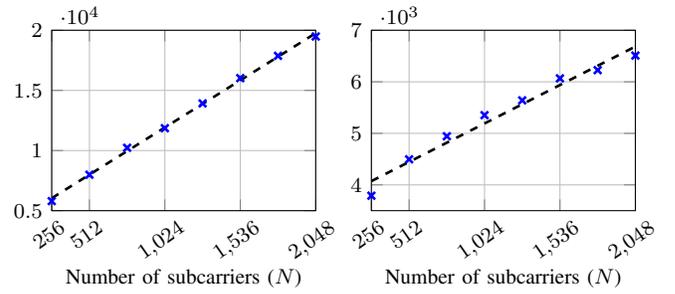

We first need a conjecture on the eigenvalues of the (Hermitian-Toeplitz)
matrix $\bT=\hat\beta\ii\hat\eta \bPsi(\hat\btau_{\hat\cala})
\bPsi\h(\hat\btau_{\hat\cala})$.
\begin{conjecture}
	\label{conj:eigvals}
	There exists an upper bound on the largest eigenvalue of \bT that grows
	linearly with $N$, i.e.,
	\begin{align*}
		\lambda_|max|(\bT)=\calo(N).
	\end{align*}
\end{conjecture}
To justify this conjecture we refer to Fig. \ref{fig:saveeig}, where the
largest eigenvalue is shown for varying $N$.

We also need a number of lemmas.
\begin{lemma}
	\label{lem:xbounds}
	There exists constants $c_1>0$ and $c_2<\infty$, such that $c_1\le
	\left<|x_i|^2\right>_{\bx_{\cald}}\le c_2$ for all $i\in\cald\cup\calp$.
\end{lemma}
\begin{IEEEproof}
	Observe that the data and pilot modulation symbol alphabets $\bba_|D|$ and
	$\bba_|P|$ only contain finite, non-zero values. We can thus take
	$c_1=\min_{x\in\bba_|P|\cup\bba_|D|}|x|^2$ and
	$c_2=\max_{x\in\bba_|P|\cup\bba_|D|}|x|^2$ to complete the proof.
\end{IEEEproof}
\begin{lemma}
	\label{lem:eigbounds}
	Assume that Conjecture \ref{conj:eigvals} holds.
	The largest and smallest eigenvalues of \bC obey
	\begin{align*}
		\lambda_|max|(\bC) = \calo(N), \qquad
		\lambda_|min|(\bC) \ge c_1\ii.
	\end{align*}
\end{lemma}
\begin{IEEEproof}
	By the Weyl inequality for Hermitian matrices $\bC$, \bT and
	$\left<\bX\h\bX\right>_{\bx_{\cald}}\ii$ we have
	\begin{align*}
		\lambda_|max|(\bC) &\le
		\lambda_|max|\!\left( \left<\bX\h\bX\right>_{\bx_{\cald}}\ii \right)
		+ \lambda_|max|(\bT).
	\end{align*}
	The first inequality follows directly from Conjecture \ref{conj:eigvals}
	and Lemma \ref{lem:xbounds}.

	Similarly by the dual Weyl inequality
	\begin{align*}
		\lambda_|min|(\bC) \ge
		\lambda_|min|\!\left(\left<\bX\h\bX\right>_{\bx_{\cald}}\ii\right)
		+ \lambda_|min|(\bT).
	\end{align*}
	Since $\hat L<N$, the matrix \bT is singular and $\lambda_|min|(\bT)=0$.
	The second inequality now follows from Lemma \ref{lem:xbounds}.
\end{IEEEproof}

By Theorem 2.2 in \cite{ng-iterative} the number of iterations required by the
CG method to achieve a desired accuracy in the solution of $\ba=\bC\bz$ is
$\calo\!\left( \sqrt{\frac{\lambda_|max|(\bC)}{\lambda_|min|(\bC)}} \right)$.
By Lemma \ref{lem:eigbounds} the number of iterations is thus $\calo(\sqrt{N})$.
Each iteration has time complexity $\calo(\hat LN)$ and the overall complexity
of solving \eqref{naive-bmu-update} via this method is therefore $\calo(\hat
LN\sqrt{N})$.

\bibliography{latex/IEEEabrv,latex/refs}

\end{document}